# HYDRODYNAMIC SIMULATIONS OF THE GROWTH OF COSMOLOGICAL STRUCTURE: SUMMARY AND COMPARISONS AMONG SCENARIOS

Jeremiah P. Ostriker & Renyue Cen

*Princeton University Observatory*

*Princeton, NJ 08544 USA*

email: cen@astro.princeton.edu

# ABSTRACT

We compute, including a current state-of-the-art treatment of hydrodynamical processes, heating and cooling, a variety of cosmological models into the extreme nonlinear phase to enable comparisons with observations.

First, we note the common, model independent results. All have a mean ($z = 0$) temperature of $10^{4.5} - 10^{5.5}$ K, set essentially by photoheating processes. Most gas is in one of two components: either at the photoheating floor $10^{4.5}$ K and primarily in low density regions or else shock heated to $10^5 - 10^6$ K and in regions of moderate overdensity (in caustics and near groups and clusters). It presents a major observational challenge to observationally detect this second, abundant component as it is neither an efficient radiator nor absorber. About 2% to 10% of the baryons cool and collapse into galaxies forming on caustics and migrating to clusters. About 1%-2% of baryons are in the very hot X-ray emitting gas near cluster cores, in good agreement with observations. These correspondances between the simulations and the real world imply that there is some significant truth to the underlying standard scenarios for the growth of structure.

The differences among model predictions may help us find the path to the correct model. For COBE normalized models the most relevant differences concern epoch of structure formation. In the open variants having $\Omega = 0.3$, with or without a cosmological constant, structure formation on galactic scales is well advanced at redshift z=5, and reionization occurs early. But if observations require models for which most galaxy formation occurs more recently than $z = 2$, then the flat $\Omega = 1$ models are to be preferred. The velocity dispersion on the $1h^{-1}$Mpc scale also provides a strong discriminant with, as expected, the $\Omega = 1$ models giving a much higher (perhaps too high) a value for that statistic.

Cosmology: large-scale structure of Universe – cosmology: theory – galaxies: clustering – galaxies: formation – hydrodynamics

# 1. INTRODUCTION

In a series of papers we have explored, with a given methodology, set of physical assumptions and numerical resolution many of the currently popular scenarios for the growth of structure in cosmology. What have we learned from this process?

Both the questions asked and the answers obtained are a strong function of the cosmic scales explored. At the small scale end ($\Delta r < 50$kpc) one can address the details of the formation of galaxies, merging of galaxies etc. To treat issues such as rotation curve morphology, disc/halo evolution, globular cluster formation etc, very high spatial resolution simulations are required for which, at the present time, the SPH method and its variants are best suited, although in the future adaptive mesh Eulerian codes or Lagrangian mesh codes may supplement the SPH approach (*cf.* Gnedin 1995; Norman & Neeman 1995). Recent examples (no attempt at completeness will be made) of such studies made at small scale are found in the work of Katz, Hernquist & Weinberg 1992; Babul & Katz 1993; Navarro, Frenk & White 1994; Summers, Davis & Evrard 1995; Steinmetz & Muller 1995). On these scales density fluctuations become very large, optical depth effects cannot be ignored and the heating and/or cooling of the gas must be treated with very great care to obtain results which are correct even to order of magnitude. To date much of the work done in this area has been based on a commendable desire for simplification, and so it has neglected fundamental processes such as photoheating and photoionization which are certain to be important given the observed background radiation fields and also has often neglected the much more difficult to quantify feedback effects consequent to star formation.

On the largest scales for which hydrodynamic calculations are relevant one needs volumes of at least $(80-120h^{-1}Mpc)^3$ to obtain representative results for the disposition of hot gas in the universe. Most of the energy density in the common scenarios is in long wavelengths, due to the fact that power spectra rise steeper than $k^{-1}$ on small scales.

Thus, the largest nonlinear waves ($\lambda \sim 50h^{-1}$Mpc) are most relevant and a somewhat bigger box is needed than this critical scale to fairly sample the nonlinear structures. In order to determine the X-ray bremsstrahlung emissivity of clusters of galaxies, the S-Z and other secondary fluctuations in the microwave background and similar phenomena, which are dependent on the overall thermal and kinetic energy density in large-scale structure, such large-scale simulations are needed. Here cooling/heating are relatively unimportant, so an adiabatic treatment is adequate, but again very high resolution is needed. A minimal number of volume elements per box of $(\sim 10^{2.5})^3 = 10^{7.5}$ is required to bridge the range from the box size needed to capture the large-scale structure to the small scale resolution needed to correctly model the density fluctuaions. Here SPH codes (Evrard 1988; Hernquist & Katz 1989; Navarro, Frenk & White 1994) fare best at the small scale end and Eulerian codes (Cen 1992; Ryu *et al.* 1993; Bryan *et al.* 1995) at the large-scale end. Both approaches are still inadequate in some essential ways and other approximate methods which combine different scales (Kaiser 1986; Bond 1990; Frenk *et al.* 1990; Bond & Myers 1991; Blanchard *et al.* 1992) are of less certain accuracy.

It is the intermediate scale, $50h^{-1}$kpc — $50h^{-1}$Mpc, which we will address primarily in this paper. The physical modelling involves elements of both regimes mentioned above but in a less extreme form. The questions asked concern the typical volume elements of the intergalactic medium (IGM hereafter), the origin of galaxies with their spatial and velocity distributions and other related issues. In a series of papers we have developed a standardized approach to the analysis of hydrodynamic phenomena on these intermediate scales. In most of the work we have ignored the feedback due to star formation. The methodology, physical assumptions, and numerical tests are presented in Cen (1992). Here we reiterate some of the essential features.

We solve simultaneously two sets of equations, one for the baryonic fluid and the other for collisionless dark matter particles. For the baryonic fluid there are eight time dependent equations as follows: the mass conservation equation of total baryonic matter, the three ionization rate equations of H I, He I and He II, the three momentum equations in three directions and the energy equation. In most of the quoted work the momentum equation (Euler's equation) is solved using a modification of the Jameson (1989) aerospace code with gravity treated using an FFT approach to the solution of Poisson's equation with periodic boundary conditions. In some of the cited papers we have used the higher resolution TVD code (Ryu *et al.* 1993) for solving the momentum equation. Locally, we also satisfy charge conservation and the gas equation of state: $P = n_{tot}kT$. The set of equations for the collisionless dark matter particles consists of three equations for change of momentum and three for change of position. In addition, we have the equation relating the density field to the gravitational forces, i.e., Poisson's equation for the perturbed density, and the two Einstein equations for the evolution of the cosmic comoving frame. The UV/X-ray radiation field (as a function of frequency and time) is calculated in a spatially averaged fashion. Changes in other quantities are computed each time step in each cell. Ionization, heating and cooling, are computed in a detailed non-LTE fashion; we allow for all the ionization states of the primary elements (H, He), computing the changes of abundance of each subspecies in each cell at each timestep, with allowance for all of the relevant atom-atom and atom-photon porcesses. Bremsstrahlung and Compton processes are most important for gas with $T > 10^{5.5}$ K , with collisional and photoionization processes most important for gas having $T < 10^{5.5}$ K. Dark matter is included in the gravitating mass with its distribution determined by the direct integration of the equation of motion and assignment from particles to cells (vice versa) using the Particle-Mesh (PM) Cloud-In-Cell (CIC) algorithm [*cf.* Hockney & Eastwood (1981); Efstathiou *et al.* (1985)]. The gravitational potential, due to both baryons and dark matter, is calculated by solving Poisson's equation with periodic boundary conditions utilizing an efficient FFT algorithm.

We have looked at four standard $\Omega = 1$, Gaussian scenarios: plain "vanilla favored" Cold Dark Matter (CDM hereafter, CO92a), tilted CDM (TCDM hereafter, CO93a),

Hot Dark Matter (HDM hereafter, CO92b), and Mixed Dark Matter (MDM hereafter, CO94a). Using the same code and set of physical assumptions we examined one open but flat model (CDM+$\Lambda$ hereafter, CGO), two non-Gaussian $\Omega = 1$ scenarios based on topological texture singularities (CDM+T and HDM+T hereafter, COST) and one open model containing only baryonic matter (PBI hereafter, COP). All eight models were studied at $128^3$ resolution (cells and particles) and with box sizes $L = (64, 16, 4, 1)h^{-1}$Mpc and corresponding nominal resolution of $(500, 125, 31, 7.8)h^{-1}$kpc at the cell size. For some of the models, smaller boxes were not computed (see column 7 of Table 1 for computed boxes). In all cases normalization was either to the first year COBE DMR $10^o$ fluctuations [CDM+$\Lambda$ ($\sigma_8 = 0.67$), TCDM ($\sigma_8 = 0.50$), MDM ($\sigma_8 = 0.67$), PBI ($\sigma_8 = 0.8$)] or to a lower level, designed to best show off the model features and hopefully not too far from correct normalization [CDM ($\sigma_8 = 0.67$), HDM ($\sigma_8 = 1.0$), CDM+T ($\sigma_8 = 0.67$), HDM+T ($\sigma_8 = 0.67$)]. Here $\sigma_8$ is the rms level of mass fluctuaions $(\Delta M/M)_{rms}$ in an $8h^{-1}$Mpc top-hat sphere at $z = 0$ as determined by linear perturbation theory. These and the other defining properties of the eight models are summarized in Table (1). Notation chosen is standard.

In addition, we studied at higher resolution ($200^3$ particles and cells) one CDM model (CDM+GF hereafter, CO92c, CO93b,c), where we allowed for feedback from star formation and followed, with a collisionless code, the galaxy subunits after they were formed. Then, after grouping these into "galaxies" we could address large-scale structure issues and galaxy properties as a function of (mean) epoch of the formation of a specific galaxy.

In Table 1 and in the subsequent discussion we group the models considered into three subsets. The first three are (biased) standard CDM models, with normalization chosen to best fit the moderate scales we are studying. They are not COBE normalized. Model (1) is the standard. The differences between models (3) and (1) should show the effects of feedback at these scales. The differences between models (2) and (1) should illustrate the effects of non-Gaussianity. The next set of four models are COBE normalized and are chosen to be representative of the surviving model types which we consider to be the remaining contenders for the title "correct". They include the tilted, mixed and lower $\Omega$ variants of the CDM scenario and a baryonic model. The final two models illustrate the HDM scenario with a favorable (non-COBE) normalization adopted – one gaussian and the other texture non-Gaussian.

In all cases we took for the gaseous baryon density at the begining of calculations the values indicated from light element nucleosynthesis by Walker *et al.* (1990) for the assumed value of $H_0$ except in the CDM+T model where we (inadvertantly) took a value slightly (17%) too low and in the PBI case, where our gaseous component is at the light element nucleosynthesis value but we allow for another collisionless component that is nominally made of material, which was originally baryonic (*cf.* Gnedin, Ostriker, & Rees 1995), and contributes approximately three times as much mass density as the gaseous component.

Despite the fact that these simulations approach the limits of technical feasibility, we are well aware that they suffer from numerical limitations at both ends of the length/mass scale. Longer waves than those included in the box would (primarily) have had the effect, had they been included, of heating the matter at late stages more than we have allowed for and correspondingly have increased the amount of hot gas, reduced the fraction of cold gas and reduced the rate of galaxy formation. The opposite effect, of reducing the mass fluctuation amplitudes at early epochs and correspondingly reducing the rate of galaxy formation, can easily be shown to the negligible. Altogether this limitation, that our largest boxes are only $64h^{-1} - 80h^{-1}$Mpc in scale, is relatively unimportant except for computations of quantities which depend specifically on the amount and density of the highest temperature gas (e.g., keV emissivity). Far more serious is our lack of resolution at small scales. Here we must be underestimating density fluctuations, cooling and galaxy formation in our big boxes. It is for this reason that we computed also the evolution in $L = (16, 4)h^{-1}$Mpc and occasionally $1h^{-1}$Mpc boxes, to see how results converged (or did not converge) as we repeated the simulations at higher and higher resolution. We find that the nonlinear mass scales converge, as $\Delta L$ becomes small, to a value of $10^{9.5} M_\odot$, corresponding to the Jeans mass of photoheated gas. Since these small boxes *overestimate* the rate of cooling and structure formation (as they omit the longer waves which would have heated the gas), we suspect that we can reasonably estimate, from combining the results obtained on various scales, the rates of cooling and galaxy formation; only higher resolution studies, now in progress, will determine this with any security.

But in any case, the primary purpose of this paper is comparative, to compare the results expected in different scenarios on the basis of calculations made with the same input physics, computational methods and spatial resolution. The different simulations listed in Table (1) have different normalizations for their power spectra. The range $(0.5 < \sigma_8 < 1.0)$ is modest, but the simulations do not all agree either in terms of $\sigma_8$ or in terms of $P_k(model)/P_k(COBE)$. Both numbers are given in column 5. Our estimate of the correct COBE normalization is based on the computation of other workers with references given in our cited papers. In general we cited work based on the $10^o$ results as likely to be most reliable. We will present the results in the following section as is, without renormalization, but comment in the text on how the results would have been changed had they all been normalized either to the same value of $\sigma_8$ or to the COBE, DMR $10^o$ observations.

Before turning to results, let us briefly examine the input power spectra of the nine models treated. The first set (Panel 1a) show the familiar CDM spectrum for which $P_k \propto k^1$ at large scales and as $k^{-3}$ at small scales. Mass fluctuations go as $k^3 P_k$ and so are dominated weakly by a logarithmic factor at scales small compared to the peak at $\lambda(= 2\pi/k)$ of $\sim 130h^{-1}$Mpc (for $h = 0.5$). The largest velocities are generated at waves when $kP_k$ peaks or at about $30h^{-1}$Mpc.

Next let us turn to the middle panel (1b). The tilted model (4) with $n = 0.7$ has, of course, relatively less small scale (large $k$) power than standard CDM for fixed long wavelength normalization. The MDM model (5) achieves the needed diminishing of the small scale power through the hot component which cannot cluster at early times on small scales. Adopting $\Omega_\nu = 0.2$ (as is currently popular) rather than 0.3, would increase small scale power below the dotted line in Figure (1) panel (b), to perhaps too great an extent. Such a change would bring it closer to COBE normalized SCDM and would produce, perhaps too large a small scale velocity dispersion. The CDM+$\Lambda$ model [dashed line (6)] has lower small scale power because with a lower value of $\Omega h$ there are two effects which reduce the small scale power: 1) less growth of these waves and, 2) the peak in the power spectrum is shifted to a large-scale by a factor $1/\Omega$, lowering the amplitude of smaller waves. The relatively large amount of large-scale power is an advantage in producing the large-scale structure, as has been noted by many investigators (*cf.* for example, Efstathiou, Bond, & White 1992). It comes from a shift in the power spectrum to smaller wavenumber which drives from a small value of $\Gamma \equiv \Omega h$. The chosen model with $\Gamma = 0.2$ is within the range allowed by Peacock and Dodds (1994), $\Gamma = 0.25 \pm 0.05$ on the basis of observed large-scale structure in the power spectrum of galaxies. The PBI model [long dashed curve (7)] has a similar amount of large-scale power to the CDM+$\Lambda$ model, it is comparable to the other models at intermediate scales, $\sigma_8 = 0.77$, but has much more very small scale power than do any of the other models treated in this paper. The peak at $\lambda \sim 250 h^{-1}$Mpc reflects the horizon at decoupling. The steep fall off from that peak (faster than $k^{-3}$) may cause pancaking as in the HDM scenario.

Finally in panel (1c) we show the HDM power spectra. The familiar cutoff at small scales which corresponds to a mass scale of $10^{15} M_\odot$ due to Silk damping of the short wavelength neutrino fluctuations has earned this model the "top down", although, as we shall see, the distinctions between "top down" and "bottom up" are more semantic than real.

## 2. RESULTS

Let us first look at the simplest integral properties of the simulations: the volume weighted temperature $\langle T \rangle$ and the density fluctuations $\sigma_M^2 \equiv 1 + (\Delta M/M)^2$ as measured on the cell scale of the different boxes. Figure (2a-2c) shows the temperature at the three scales for the nine models and (3a-3c) the density fluctuations. We see that in general, for most models at most epochs the mean temperature is an increasing function of time. This is simply due to the fact that the scales at which waves are nonlinear increase with time rapidly enough so that the nonlinear velocity scale $v_{NL} \equiv H\lambda_{NL}(physical)$ increases. Thus the characteristic temperature will grow as well since $kT_{NL}/m_0 \approx v_{NL}^2$. In the PBI model an assumed early generation of quasars produces a large high energy background which is effective in Compton heating the gas. For all of the viable models (the second set), the final temperatures due to collapsing structures are similar, $T \sim 10^{4.7}$ K for the 500kpc cell, $10^{4.1}$ K for the 125kpc cell and $10^{3.2}$ K for the 31kpc cell. The reason for the dependence on averaging scale is that with increasing resolution, one reaches higher densities (*cf.* Figures 3a-c) and correspondingly more efficient cooling. Temperatures due to shocks from structure formation start to increase earlier in models (2) and (6) because structure itself starts earlier in non-Gaussian and in lower $\Omega$ variant models. The burst of galaxy formation, which occurs in model (3) (which allows for this feedback from supernovae), accounts for the rapid rise in the temperature at redshift $4-5$ in this model. In the HDM models (not studied, for obvious reasons, in small boxes) very little happens till very late times ($z \sim 2-1$), when the large waves begin to break.

Now we turn to the density fluctuations portrayed in Figures 3a-c. In panel (a) we see that early growth of structure on the 500kpc scale is largest, as expected, in the open or non-Gaussian models (2,6,7) and the final amplitude largest as expected in the PBI model (7), which has the most small scale power. This should provide an important discriminant for this model. On the smaller $125h^{-1}$kpc scale the open models have largest amplitudes, especially at early times but again the non-Gaussian model (3) achieves early structure formation. On the smallest scale the amplitudes are relatively large, with standard CDM the largest (of those shown); the reason that model (3) has much less structure in the gaseous fluid is that very high density lumps on small scales have collapsed to form galaxies. If we compare the dark matter density fluctuations in model (1) and (3) on this scale, the situation is inverted with the largest fluctuations in model (3) larger than in model (1); the dark matter has been pulled into tighter knots by the cooling and collapsing baryonic fluid.

We have found it useful to subdivide the gas into four components which make a complete but crude classification possible: (1) virialized, bound, hot objects, which on the large scales represent the gas in clusters of galaxies and on the small scales represent the $L_\alpha$ clouds — "Virialized Gas"; (2) bound, cooled objects, i.e., collapsed compact objects — "Galaxies"; (3) unbound, hot regions with temperature $\geq 10^5$ K — "Hot IGM"; (4) other regions, primarily — "Voids". The break point at $10^5$ K is adopted because it is past the peak of the "cooling curve". We have tabulated for each of our nine models the mass and volume fraction for each component at a variety of redshifts and levels of resolution. The results for the mass weighted fraction are collected in Figures (4a-c) "Virialized Gas", (5a-c) "Galaxies", (6a-c) "Hot IGM" and (7a-c) "Voids" (shown as 1-mass fraction for more convenient display).

In Figure 4 we see that for most models on most scales the mass fraction in the virialized (hot) gas component reaches about $f \approx 10^{-3\pm1}$, with largest values in the lower $\Omega$ models and smallest for standard CDM. As expected, the non-Gaussian model and the low density models reach this level first as they reach strong nonlinearities first. The most significant and dramatic differences amongst the models are shown in Figure (5a), where indicative "galaxy" mass fraction is shown. Only for model (3) is this calculated correctly as collapsing gas lumps are changed irreversibly into "stellar" collisionless particles in that integration. We see that in the more approximate calculation (1) using the same initial conditions, aside from the turn down (as the galaxies are "evaporated") at late epochs, models (1) and (3) are reassuringly similar, indicating that our very rough treatment is not seriously in error. What we see is that about 1% if the baryon mass typically will collapse into galaxies. In the higher resolution, smaller boxes the fraction reaches about 2% of the total baryons. In Figure (5a) the middle panel shows how the currently viable models differ from standard ($b = 1.5$) CDM. In the two low density models galaxies form at the same time or earlier, but in the tilted or mixed models galaxy formation is quite late. In 500kpc cells galaxy formation in these models starts only at $z = 2.5$ and even in the 31kpc cell run (Figure 5c) initiation occurs at $z = 3.5$, which would lead to difficulies in satisfying the Gunn-Peterson test for high redshift quasars. For PBI the potential problem is the opposite, galaxy formation might occur at such an early epoch as to possibly make galaxies much more dense than real galaxies are observed to be. Higher resolution studies would be required to address these problems.

It is worthwhile to understand why in our simulations the "overcooling problem" (see White & Frenk 1991; Kauffmann, White, & Guiderdoni 1993; Lacey *et al.* 1993) does not occur, while some early studies of galaxy formation found that small galaxies at high redshift accreted a large fraction of the gas. The effect of photoionization on slowing or reversing the cooling rate and hence stablizing the collapse on small scales has recently been investigated (Efstathiou 1992; Quinn, Katz, & Efstathiou 1995; Steinmetz 1995). Since our smallest boxes having resolution of $7.8h^{-1}$kpc, adequate to resolve smallest structures determined by Jeans' instability, do not show a large fraction of collapsed baryons, it seems that the photoionization effect, which is included in all the simulations presented here, is the major effect reversing the overcooling problem. Furthermore, we know that the collapsed baryon fraction is already overestimated in the small boxes because of the absence of long waves (longer than the box size), whose breaking would have heated gas to a higher temperature and hence reduce the cold fraction. Our new simulations with even higher effective resolution (also with photoionization effect included) yield results of the collapsed fraction consistent with the results presented here (Miralda-Escudé et al. 1995).

Most of the mass of the universe resides in the two remaining categories: Hot IGM, shock heated, unbound hot gas around clusters and groups and cool gas in the voids. These two phases dominate at late epochs, with comparable mass in the two components and only small differences among the models. The volume fractions (not shown) are not as different as one might expect, the volume fraction of void gas being somewhat larger than its mass fraction, due to having lower mean density. These two phases are far from being in mechanical equilibrium. The cool gas in the voids, to the extent that it contains small scale power, is identified, observationally with the Lyman alpha forest (Cen et al. 1994; Miralda-Escudé et al. 1995; Hernquist, Katz, & Weinberg 1995; Zhang, Anninos, & Norman 1995).

*But the "Hot IGM" component which has a comparable mean mass density has not been identified observationally.* The temperature range is occupied, $10^5 - 10^7$ K, makes it difficult to observe as it emits in the soft X-ray bands where the Galactic component of the observed radiation field is large. In Cen et al. (1995) we argue that the diffuse, soft X-ray cosmic background identified by the ROSAT observatory (Hasinger et al. 1993) is exactly this component.

Figure (8) shows the final ($z = 0$) mean radiation fields in the various models. All allow for (box averaged) line absorption, recombination and bremsstrahlung emission. Model (3) also includes stellar UV output following the prescription of Scalo (1986). At 1keV the COBE normalized models produce about $10-20\%$ of the soft X-ray background. This is perhaps 1/3-1/2 from identifiable X-ray emitting clusters ("Hot virialized gas") and perhaps 2/3-1/2 from background bremsstrahlung gas emission (*cf.* Cen et al. 1995). Of course most ($\sim 80-90\%$) of the X-ray background at 1keV is due to a discrete set of sources associated with AGN.

The spectrum in the range 0.5-1.0keV is steeper than the total XRB, implying that hot gas makes a relatively larger contribution at lower energies. Between 13.6eV and about 300eV the absorption edges of cosmologically distributed H and He will absorb most of the emitted radiation. Below 13.6eV only model (3) which allows for input from hot stars has a significant background radiation field. At very high energies ($> 10$keV) one model stands out as distinct. The PBI model, because of the large amount of small scale power even at the epoch of recombination, is able to convert most of its baryonic matter into compact objects at very early times. In one scenario (Gnedin & Ostriker 1994), which we adopted in our model, these compact objects are black holes with masses in the range $10^6 - 10^8 M_\odot$, which subsequently act as the central engines of luminous, quasar-like systems. The immense amount of high energy radiation from these systems creates the diffuse high energy background radiation field at high redshift ($z \sim 500$). The high energy ($> 10$keV) photons at lower redshifts are the relics of this field.

In Table (2) we collect other output from the models, such as estimates for the Sunyaev-Zel'dovich $y$ parameter, and 1keV volume emissivity, the neutral hydrogen fraction, the bias $b_8$ of galaxy fluctuations over mass fluctuations in $8h^{-1}$Mpc spheres and particle-particle pairwise velocity dispersion at separation $1h^{-1}$Mpc. All the quantities are computed at redshift zero. In order to estimate the possible systematic errors on various computed quantities, especially the thermodynamic quantities, we have rerun some of models (Models #1, 6, 7) with a new, higher resolution, shock capturing TVD code (Ryu *et al.* 1993). The results from both codes are listed (Columns 2,3,4 of Model 1,6,7). We note that, while the X-ray emissivities differ by a large factor (10-30) within the same model from those with the higher resolution code, (TVD) being larger than those from the lower resolution code (CEN), as expected, the Sunyaev-Zel'dovish parameters hardly change, making it a robust statistic to compute. The neutral hydrogen fractions (column #4) differ by some intermediate factors (1.2-6.0).

Of the quantities listed in Table 2 perhaps the most interesting one is shown in the last column. The one dimensional relative velocity dispersion of galaxies separated by 1Mpc is observed to be 340±40km/s (Davis & Peebles 1983). The models can be divided into three groups. HDM based models (8 and 9) predict far too high a small scale velocity dispersion even though they have almost no initial power on the quoted small scales. This is a totally nonlinear phenomena. The $\Omega = 1$ variants of CDM (models 1-5) probably also predict too high a value of $v_{1d}$, but not by as large a factor. The models with less than closure density (6 and 7) give a much lower value of $v_{1d}$ with the PBI prediction perhaps too low. By this measure ($v_{1d}$) the low density CDM models (with or without $\Lambda$) are the "best choices".

## 5. DISCUSSION

This is not an appropriate place to rate or rank order the models in terms of astrophysical plausibility. Many considerations, other than those discussed in this paper are necessary for that exercise. The age and Hubble constant issue, gravitational lensing, the evolution of clusters etc should all be considered in addition to the factors treated here, and it would be necessary to make a quite careful assessment of the observational errors associated with each putative test. Some of these matters are addressed in Ostriker & Steinhardt (1995) where the CDM+$\Lambda$ model was found to be relatively most attrative.

Rather, we will in this section describe the "family traits" of the different types of models and leave it to the reader to determine which are most attractive in matching observational requirements.

If early structure formation is desired, then low density models fare best. Should there develop compelling evidence that the bulk of the galaxies (or even of the galactic spheroids) were formed before redshift four, then all of the Tilted and Mixed models having $\Omega_{matter} = 1$ would be firmly excluded, and our attention should turn to $\Omega_{matter} \leq$ 0.5 variants such as CDM+$\Lambda$ or PBI. Conversely, if most galaxy formation occured later than $z = 2$, then only the $\Omega_{matter} = 1$ models are plausible. The second feature to look at is scale. PBI has the scale of horizon at decoupling imprinted on its power spectrum and also has relatively more very small scale power. The first of these signatures is intrinsic to the model and not very strongly dependent on "m", the assumed power law index for perturbations. The observational signature of this power should be occasional very large-scale $\sim 100h^{-1}$Mpc pancakes, and, in addition relatively large values for the bulk flow within some $50h^{-1}$Mpc spheres.

The epoch of the reionization of the universe and associated production of the Gunn-Peterson effect (transparency to Lyman alpha radiation) and the subsequent formation and later evaporation of the Lyman alpha forest may be expected to be a strong indicator of the small scale power as well as $\Omega_{matter}$. However, the simulations summarized in this paper are not of high enough resolution nor detailed enough in treatment of the physics to make conclusive remarks, so this issue is left for further study.

The remarks made above concern the differences amongst models and the future work observationally and computationally that will be required to discriminate among them. Of equal or greater importance are the areas where all models agree in their essential predictions. Here we can hope that the numerical simulations are telling us something fairly definitive about the real world, either providing physical explanations for observed phenomena or making predictions fo new phenomena to be observed.

The mean temperature of all the viable models is in the range $10^{4.5} - 10^{5.5}$ K. The floor of about $10^{4.3}$ K is set by photoheating (for all except the very small fraction of optically thick material) with the maximum of $10^{8.3}$ K appropriate to the small fraction of gas which has fallen into rich clusters. A much larger mass fraction has been heated by shocks from supernovae or from falling into caustics/galaxy groups to a temperature of $10^5 - 10^6$ K. The large mass of gas expected in the temperature range of $10^5 - 10^7$ K is a definite prediction of those numerical simulations. Detecting it will present a challenge to observers.

The virialized hot gas at high temperatures should constitute about 1% of the total baryonic mass density. This gas is well observed as the abundant X-ray emitting gas in rich clusters of galaxies, in good agreement with the model predictions (Kang *et al.* 1994; Cen & Ostriker 1994; Bryan *et al.* 1995). This agreement between computation and observation, while not yet sufficiently detailed to conclusively discriminate *between* models, does provide strong evidence that the simulations are describing the real world.

The models predict that roughly 2% to 10% of the baryons should be expected to cool and collapse into galaxies. While these simulations can say almost nothing about the detailed properties of the predicted objects, the predicted mass fraction is moderately secure and in reassuring agreement with observed estimates of $\Omega_{gal}$: $\Omega_{baryon,gal}(predicted) \approx 0.05 \times 0.04 \approx 0.002$ or $\rho_{gal} \approx 10^9 M_\odot/\text{Mpc}^3$. This value is close to the observed light density of $1.5 \pm 0.4 \times 10^8 L_\odot h\text{Mpc}^{-3}$ (De Lapparent, Huchra & Geller 1989) times an observational baryonic mass to ligh ratio of 4. The agreement to better than order of magnitude accuracy leads to some confidence that the detailed modelling may include the most essential physical processes, i.e, we understand the rough observed cosmic mass density in galaxies.

The baryonic component in the "voids" – regions of relative underdensity – should, according to these calculations, comprise perhaps 30% of the total baryonic mass. It is kept by photoionization processes at a temperature of about $10^{4.3}$ K and a neutral fraction (in the clumps) about $10^{-5}$. As noted in Cen *et al.* (1994) the predicted properties of the gas agree quite well with broadly distributed Lyman alpha gas.

The predicted UV and X-ray background from large-scale structure formation (Figure 8) is a significant fraction of the observed background only in the range $0.5 - 1.0$keV (*cf.* Cen *et al.* 1995), and this prediction is quite testable by looking at the observed spatial correlation signal. The UV component predicted is of course dependent on what assump-tions are made for galaxy and quasar formation but with efficiencies chosen to be as observed, combined with the computed formation rates the models provide an adequate fit to both the temporal and frequency dependence of the UV background.

Overall, these correspondences between the ab initio numerical simulations and cosmologically observed quantities encourages us to believe that further more detailed work in the area will be scientifically rewarding: our corrent models for the origin of cosmic structure and our methods of computation, if not yet "correct", contain substantial elements of truth.

The work is supported in part by grants NAGW-2448, NAG5-2759, AST91-08103 and ASC93-18185. It is a pleasure to acknowledge NCSA for allowing us to use their Convex-3880 supercomputer where some of the models were computed. We would like to thank the referee Dr. A. Blanchard for a careful reading of the paper and useful comments. We would like to thank the hospitality of ITP during the galaxy formation workshop when this work was brought to near completion, and the financial support from ITP through the NSF grant PHY94-07194.

FIGURE CAPTIONS

Fig. 1– Figure (1) shows the power spectra for for the nine models. The models are broken into three groups: left panel for CDM models, middle panel for COBE normalized model, and right panel for HDM based models. Different curves are labelled correspondingly (see Table 1).

Fig. 2– Figures (2a,b,c) show the volume-weighted temperatures at the three scales (500,125,31)kpc/h, respectively. Not all models are shown in Figures (2b,c) simply because they were not computed.

Fig. 3– Figure (3) shows the corresponding density fluctuations at the three scales (500,125,31)kpc/h, respectively.

Fig. 4– Figure (4) shows the corresponding virialized mass fraction at the three scales (500,125,31)kpc/h, respectively. Note that HDM model (model 8) is not shown in Figure 4a because it predicts a value below the displayed value.

Fig. 5– Figure (5) shows the corresponding galaxy mass fraction at the three scales (500,125,31)kpc/h, respectively.

Fig. 6– Figure (6) shows the corresponding the IGM mass fraction at the three scales (500,125,31)kpc/h, respectively.

Fig. 7– Figure (7) shows the corresponding the mass fraction in voids at the three scales (500,125,31)kpc/h, respectively.

Fig. 8– Figure (8) shows the UV/X-ray radiation fields of the nine models at redshfit zero.

**Table 1.** List of parameters for the nine runs

| Model | $(\Omega, \lambda)$ | $h = H/100$ km/s/Mpc | $\Omega_b$ | Normalization $\sigma_8$/COBE | Gaussian | Boxes*** | Reference |
|---|---|---|---|---|---|---|---|
| CDM (1) | (1,0) | 0.50 | 0.060 | 0.67/0.64 | yes | (64,16,4,1) | CO92a |
| CDM+T (2) | (1,0) | 0.50 | 0.050 | 0.67/>1.00 | no | (64,16) | COST |
| CDM+GF* (3) | (1,0) | 0.50 | 0.060 | 0.77/0.73 | yes | (80,8) | CO92c,93b,c |
| TCDM (4) | (1,0) | 0.50 | 0.060 | 0.50/1.00 | yes | (64,16,4,1) | CO93a |
| MDM (5) | (1**,0) | 0.50 | 0.060 | 0.67/1.00 | yes | (64,16,4,1) | CO94 |
| CDM+$\Lambda$ (6) | (0.3,0.7) | 0.67 | 0.034 | 0.67/1.00 | yes | (64,16,4,1) | CGO |
| PBI (7) | (0.15,0) | 0.80 | 0.036 | 0.77/0.77 | yes | (64) | COP |
| HDM (8) | (1,0) | 0.75 | 0.027 | 1.00/2.00 | yes | (64) | CO92b |
| HDM+T (9) | (1,0) | 0.75 | 0.028 | 1.00/>1.00 | no | (64) | COST |

* Feedback from star formation included with $\epsilon_{UV} = 10^{-4.0}$ and $\epsilon_{SN} = 10^{-4.5}$.

** $\Omega_{cold} = 0.64$, $\Omega_{hot} = 0.3$, $\Omega_b = 0.06$.

*** Computed boxes for the model in $h^{-1}$Mpc.

**Table 2.** Summary of various computed quantities for the nine runs

| Model Run | TVD $y_0$ CEN | $\epsilon_\nu(1\text{keV})^*$ TVD/CEN | $HI/H_{tot}$ TVD/CEN | $b_{8,g}$ | $v(1d)(\text{km/s})$ |
|---|---|---|---|---|---|
| 1 | $(1.3 \pm 0.7) \times 10^{-6}$ $(1.3 \pm 0.7) \times 10^{-6}$ | 45.0/16.1 | 0.15/0.19 | 1.9 | 520 |
| 2 | $(3.4 \pm 1.7) \times 10^{-6}$ | 16.1 | $4.6 \times 10^{-5}$ | 2.0 | 530 |
| 3 | $(1.5 \pm 0.8) \times 10^{-6}$ | 128.9 | $2.5 \times 10^{-7}$ | 1.5 | 650 |
| 4 | $(5.4 \pm 2.7) \times 10^{-7}$ | 8.56 | 0.51 | 2.4 | 550 |
| 5 | $(5.4 \pm 2.7) \times 10^{-7}$ | 19.6 | 0.42 | 1.7 | 605 |
| 6 | $(3.5 \pm 1.8) \times 10^{-7}$ $(3.6 \pm 1.8) \times 10^{-7}$ | 1.85/0.15 | 0.13/0.43 | 1.5 | 380 |
| 7 | $(1.1 \pm 0.6) \times 10^{-5}$ $(1.2 \pm 0.6) \times 10^{-5}$ | 62.6/2.80 | 0.015/0.088 | 1.1 | 235 |
| 8 | $(3.2 \pm 1.6) \times 10^{-6}$ | 43.3 | 0.42 | 4.5 | 1100 |
| 9 | $(2.3 \pm 1.2) \times 10^{-6}$ | 35.7 | 0.36 | 3.2 | 1200 |

$^*$ in units of $10^{-56} \text{erg/cm}^3/\text{hz/sec/sr}$

In columns #2,3,4 some of the models contain two entries; the first entry is from the TVD code (Ryu *et al.* 1993) and the second entry from Jameson code (Cen 1992).

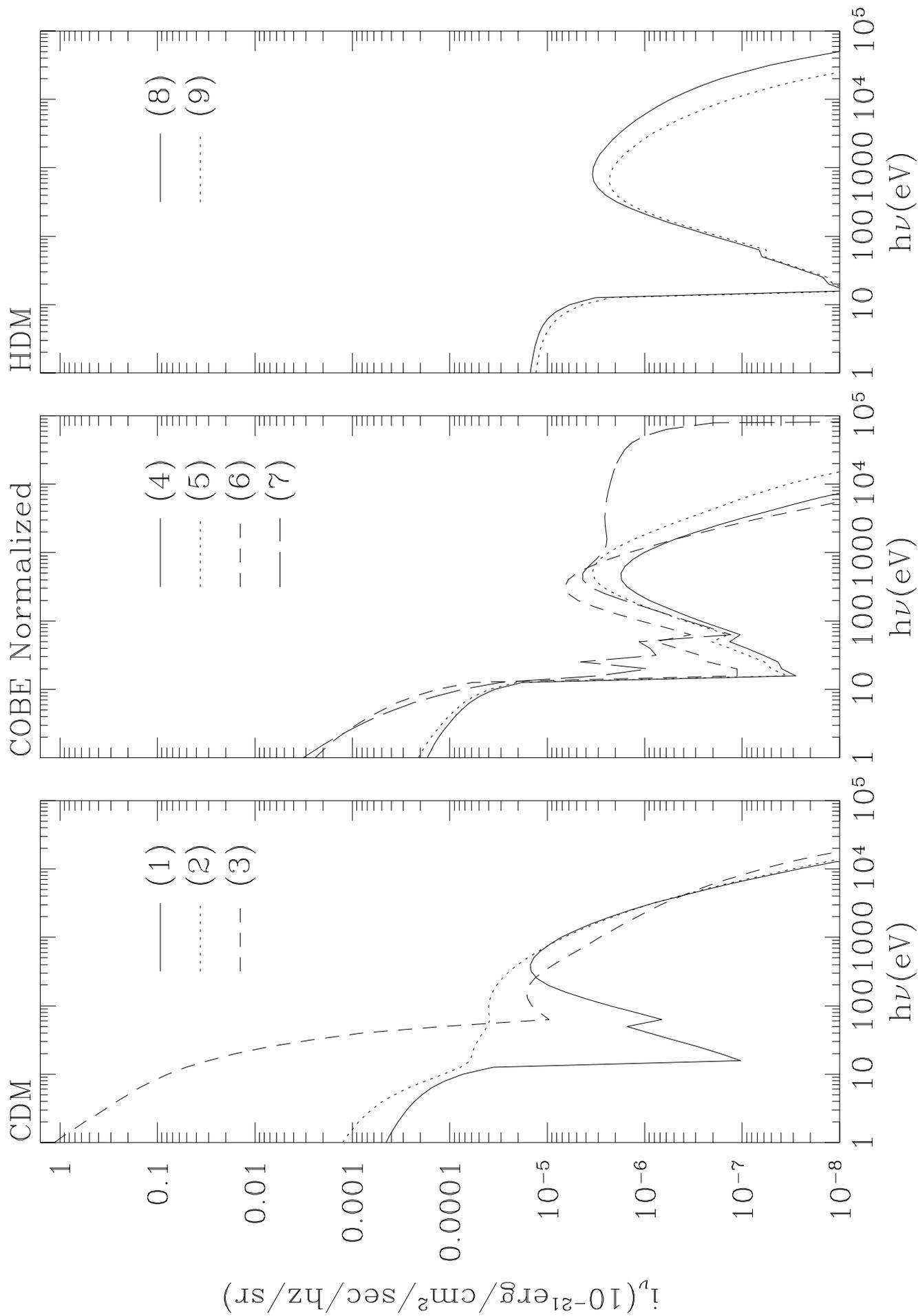

Figure 8

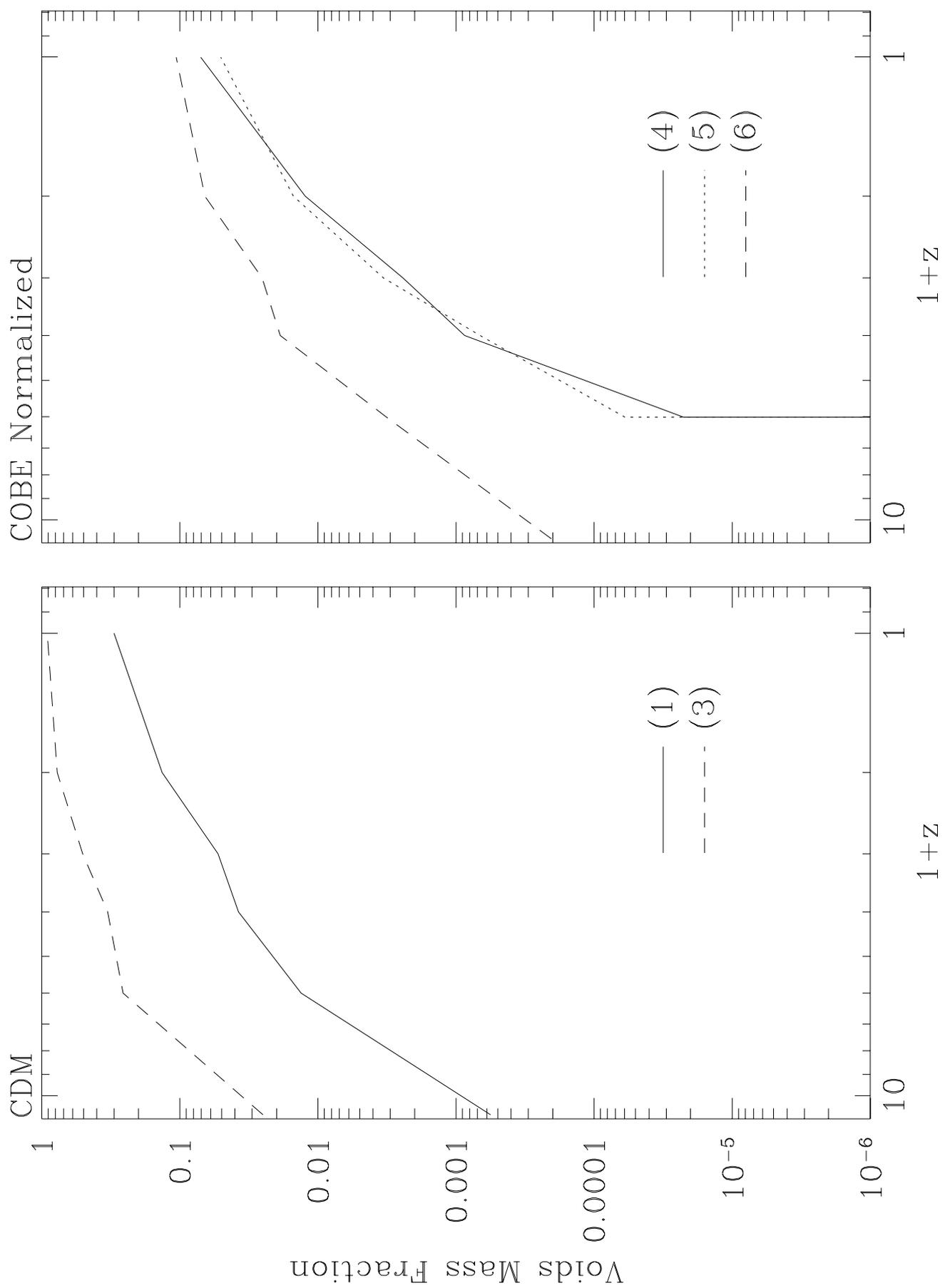

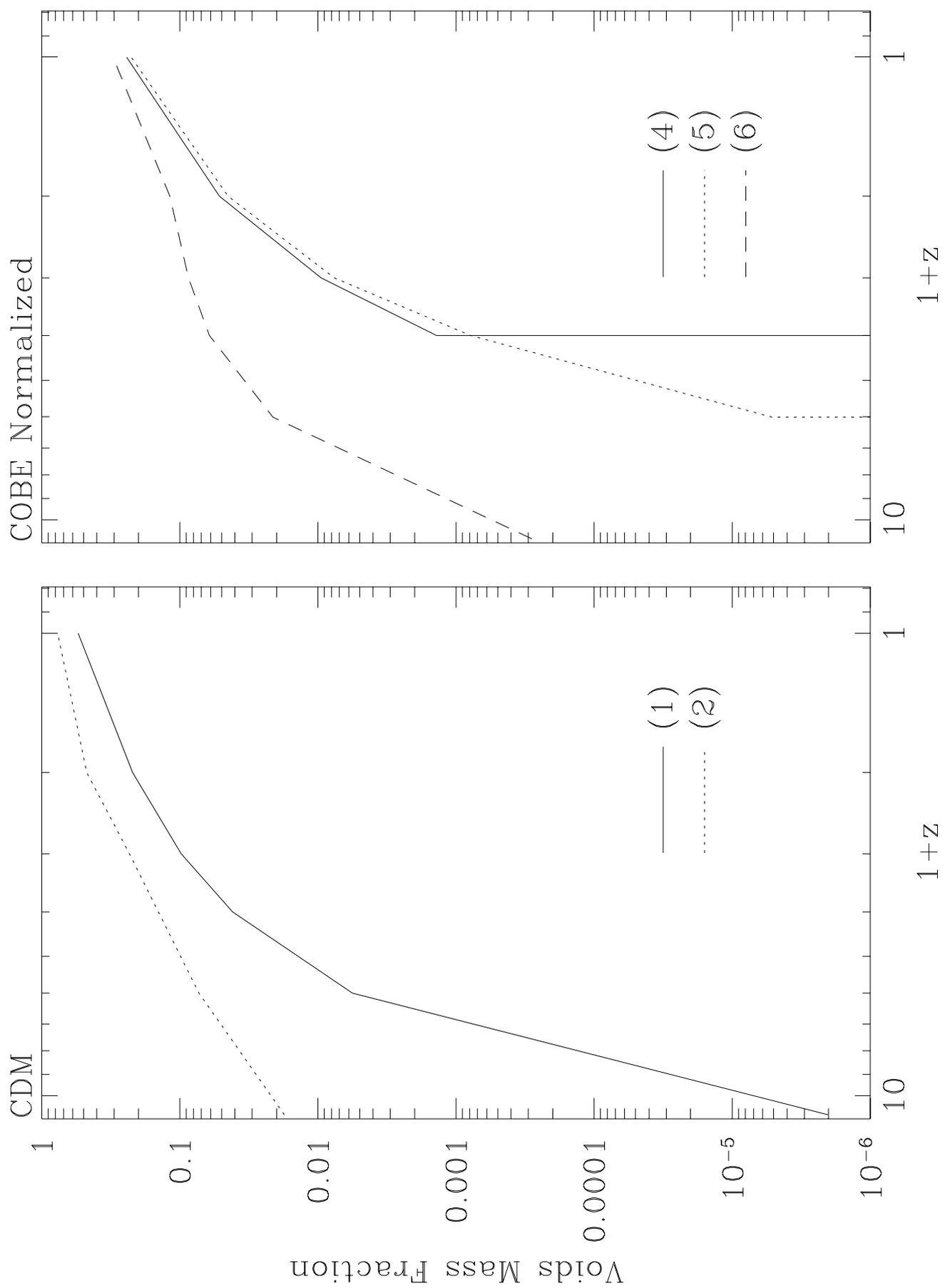

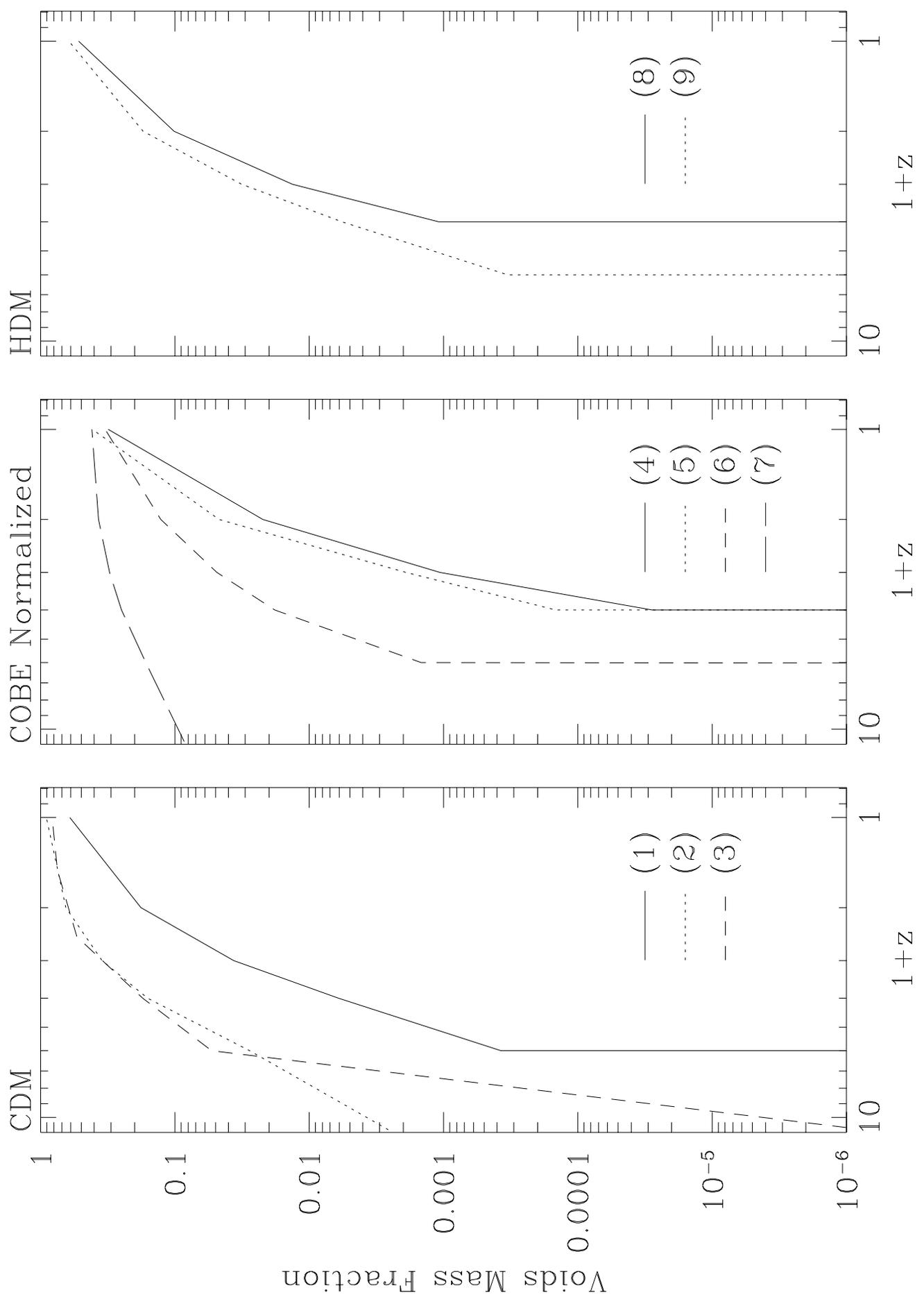

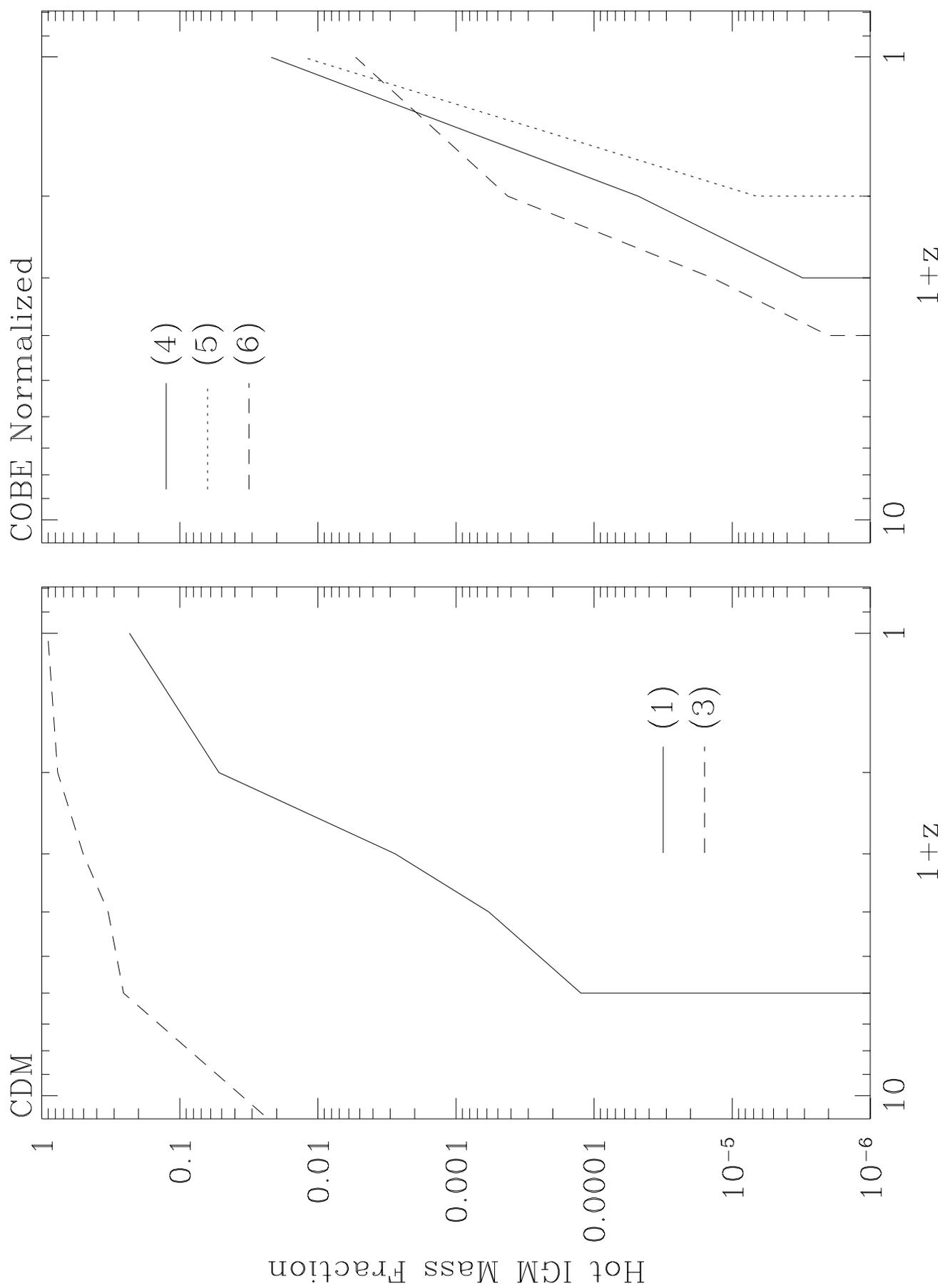

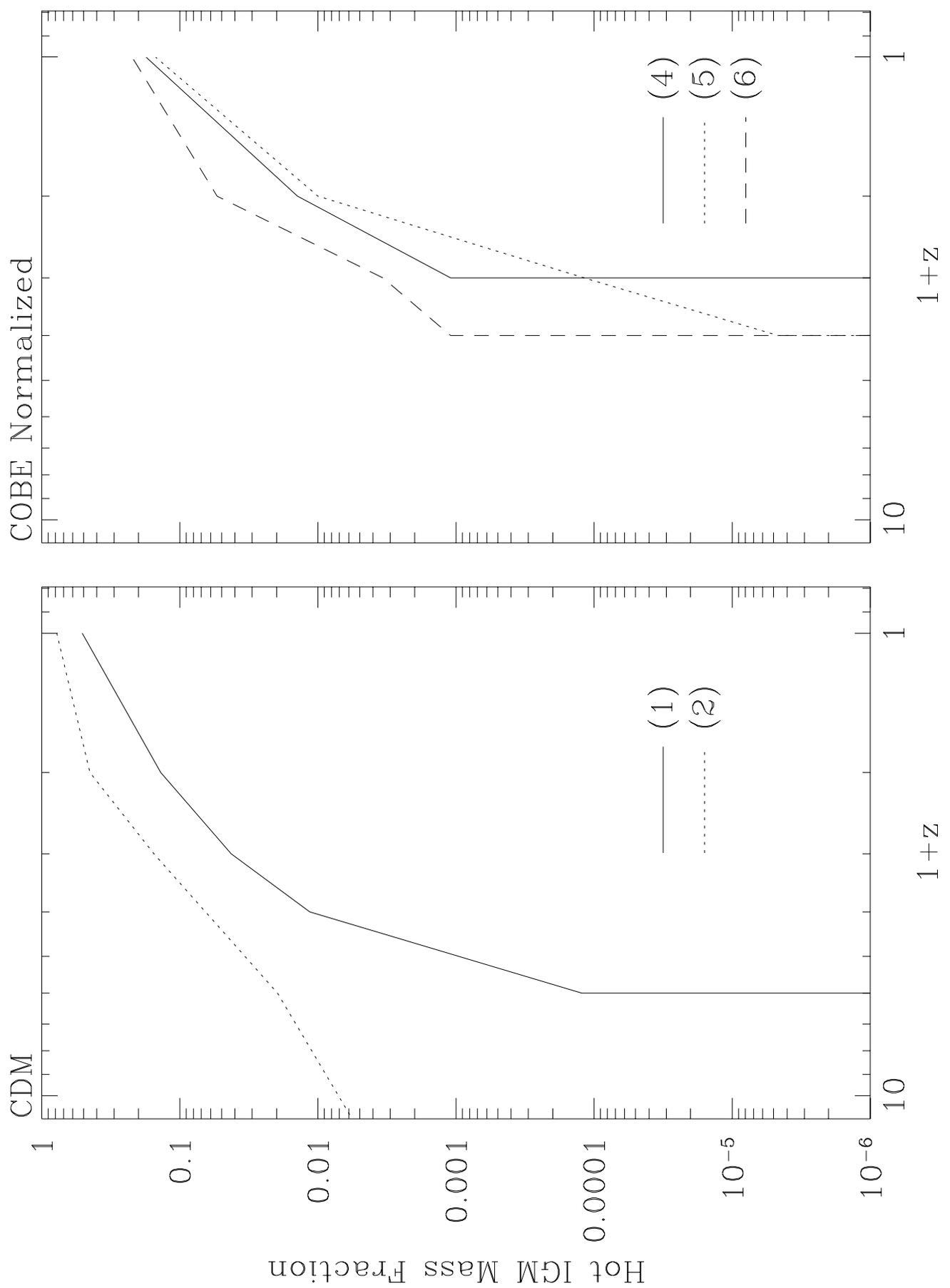
Figure 6b

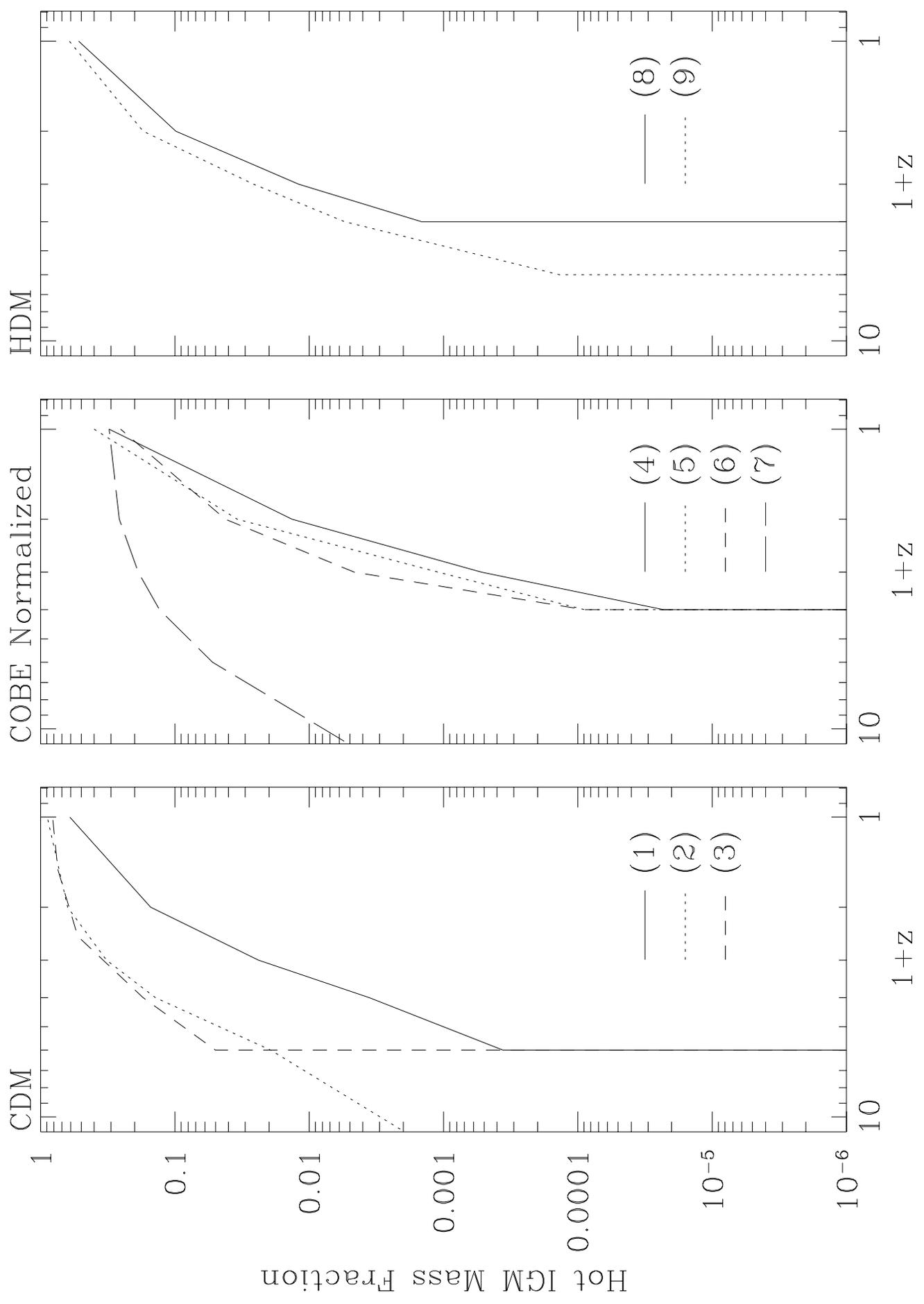

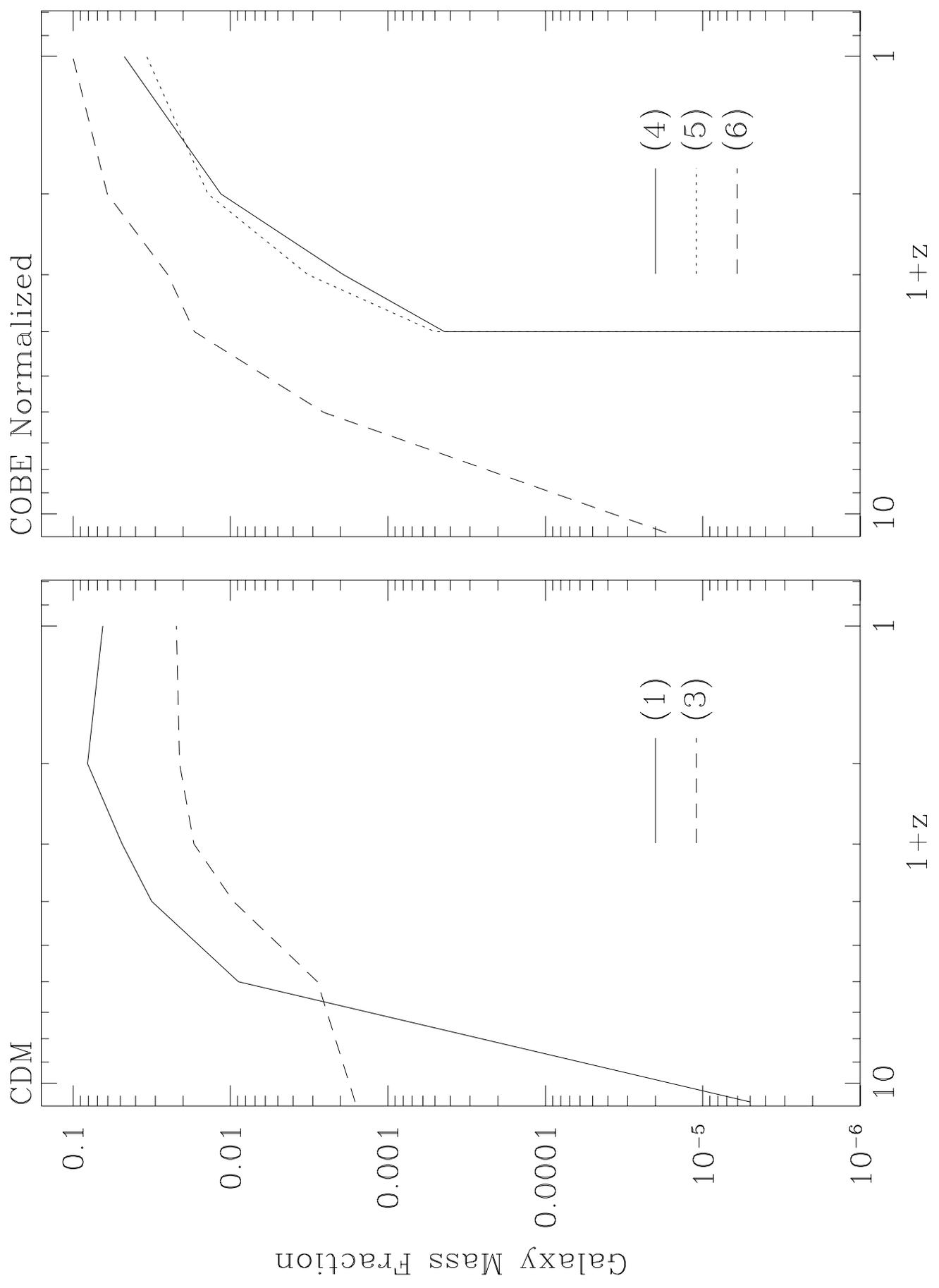

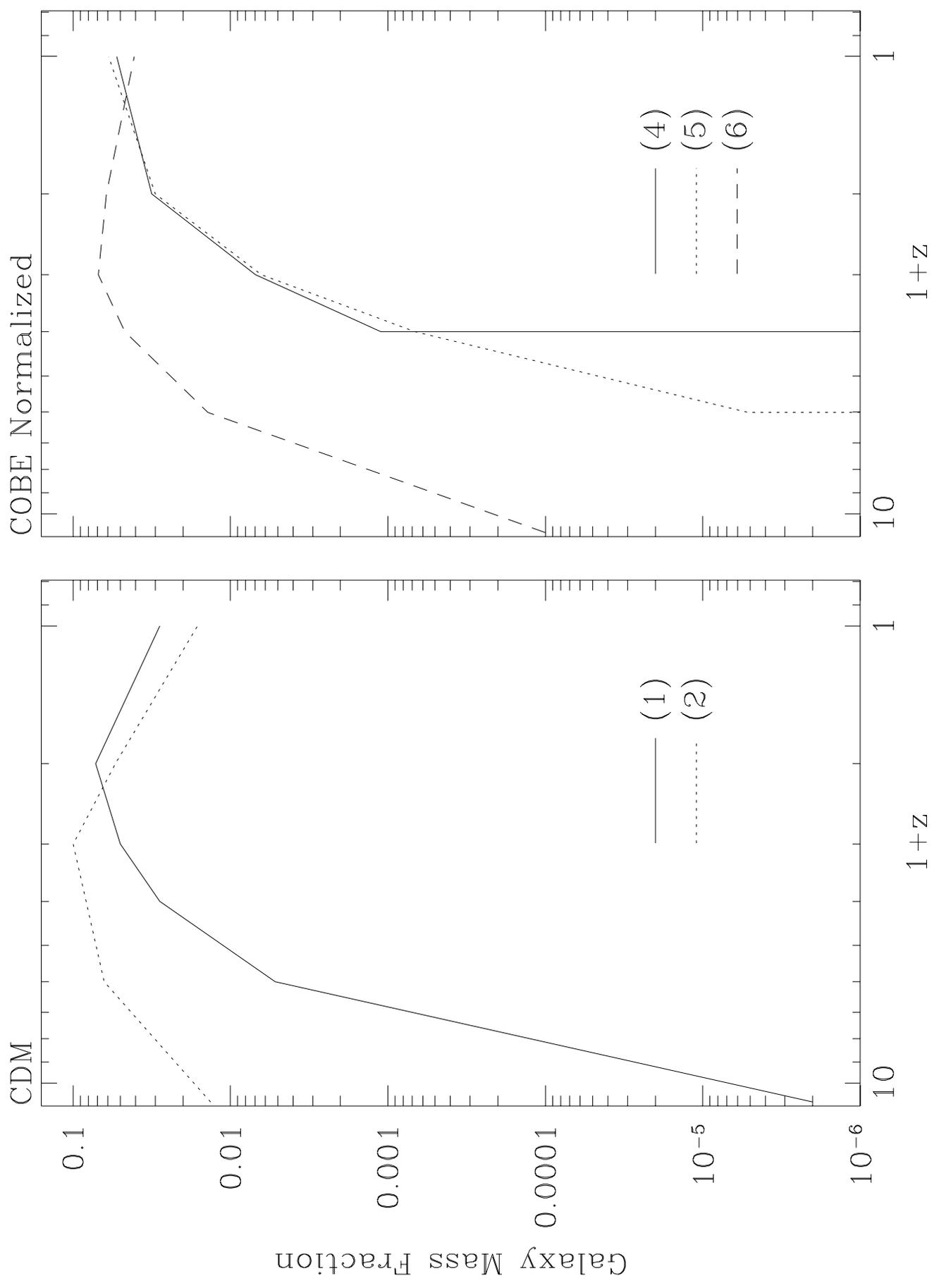

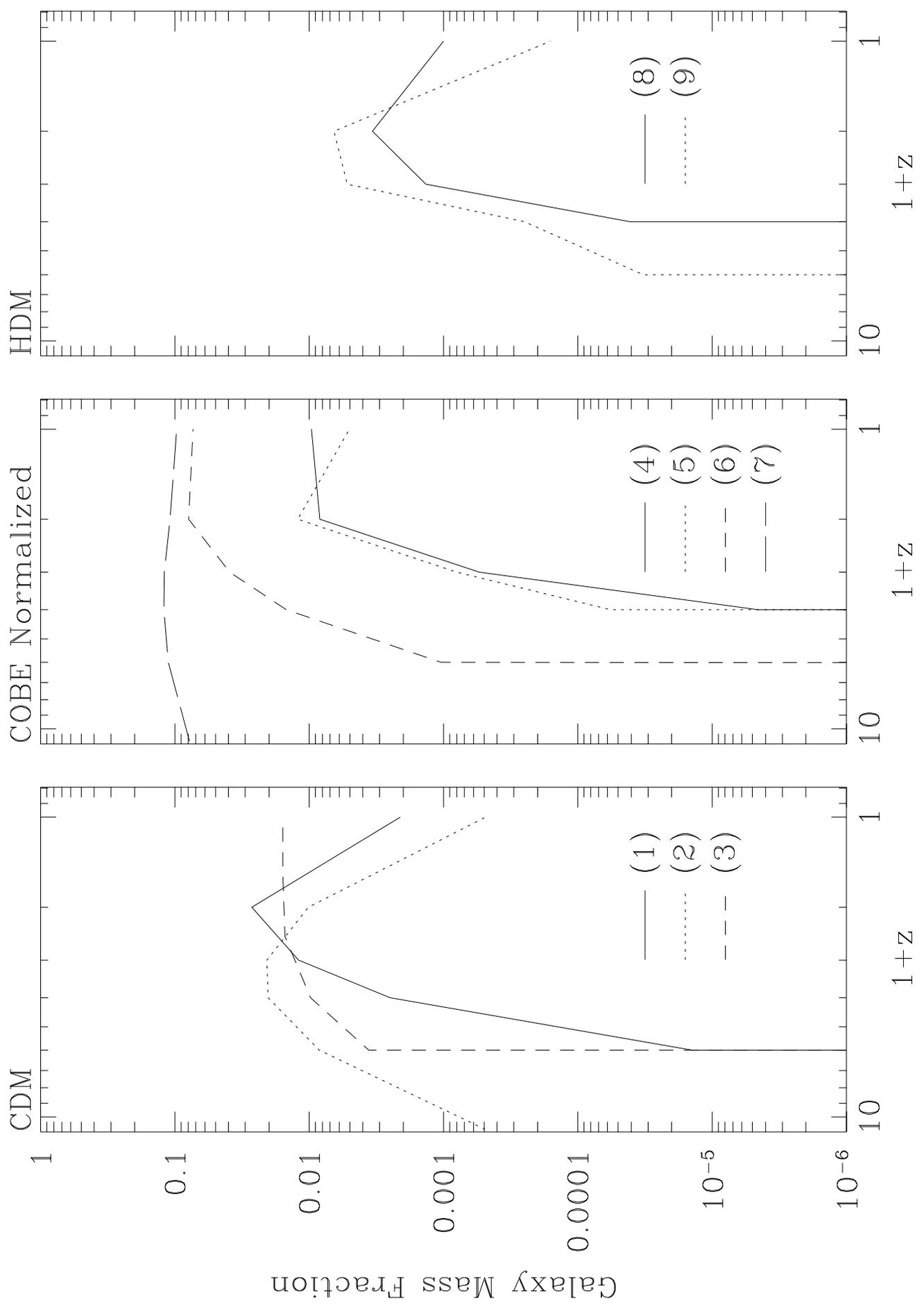

Figure 5a

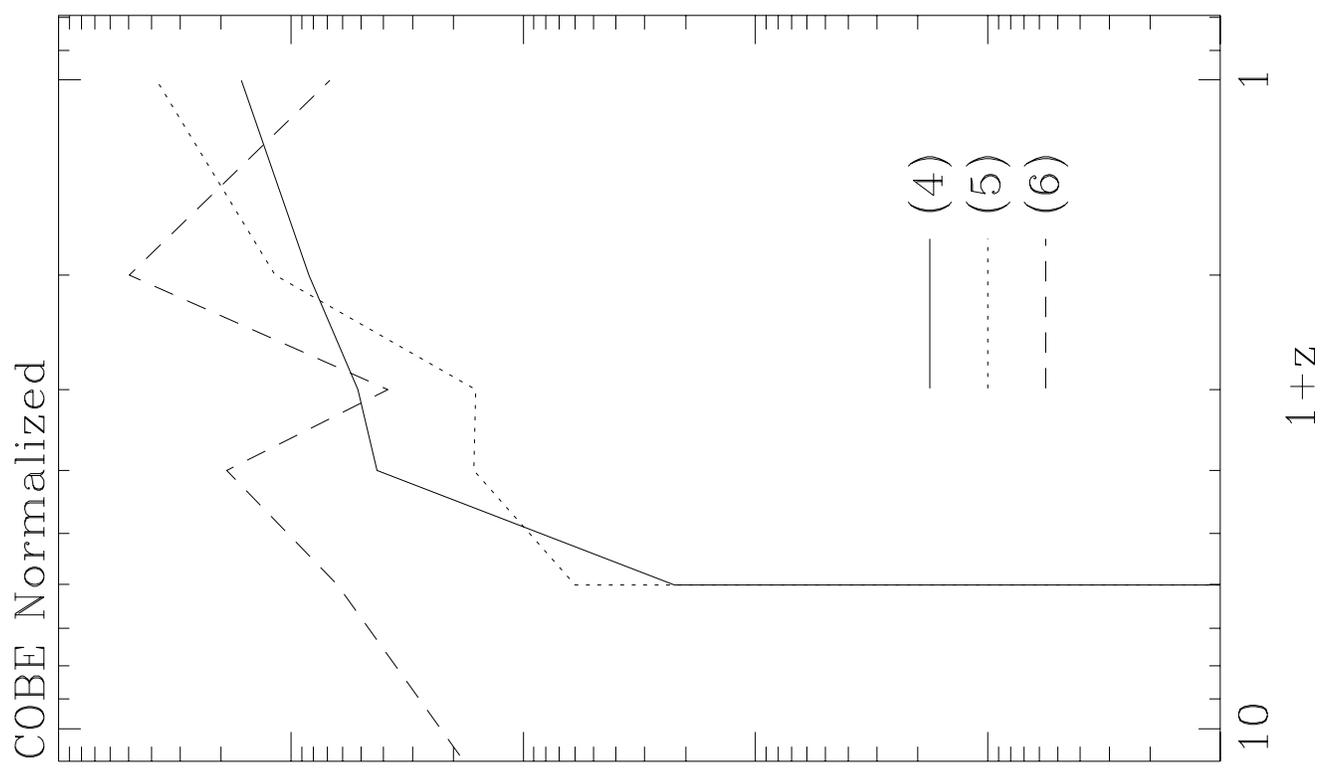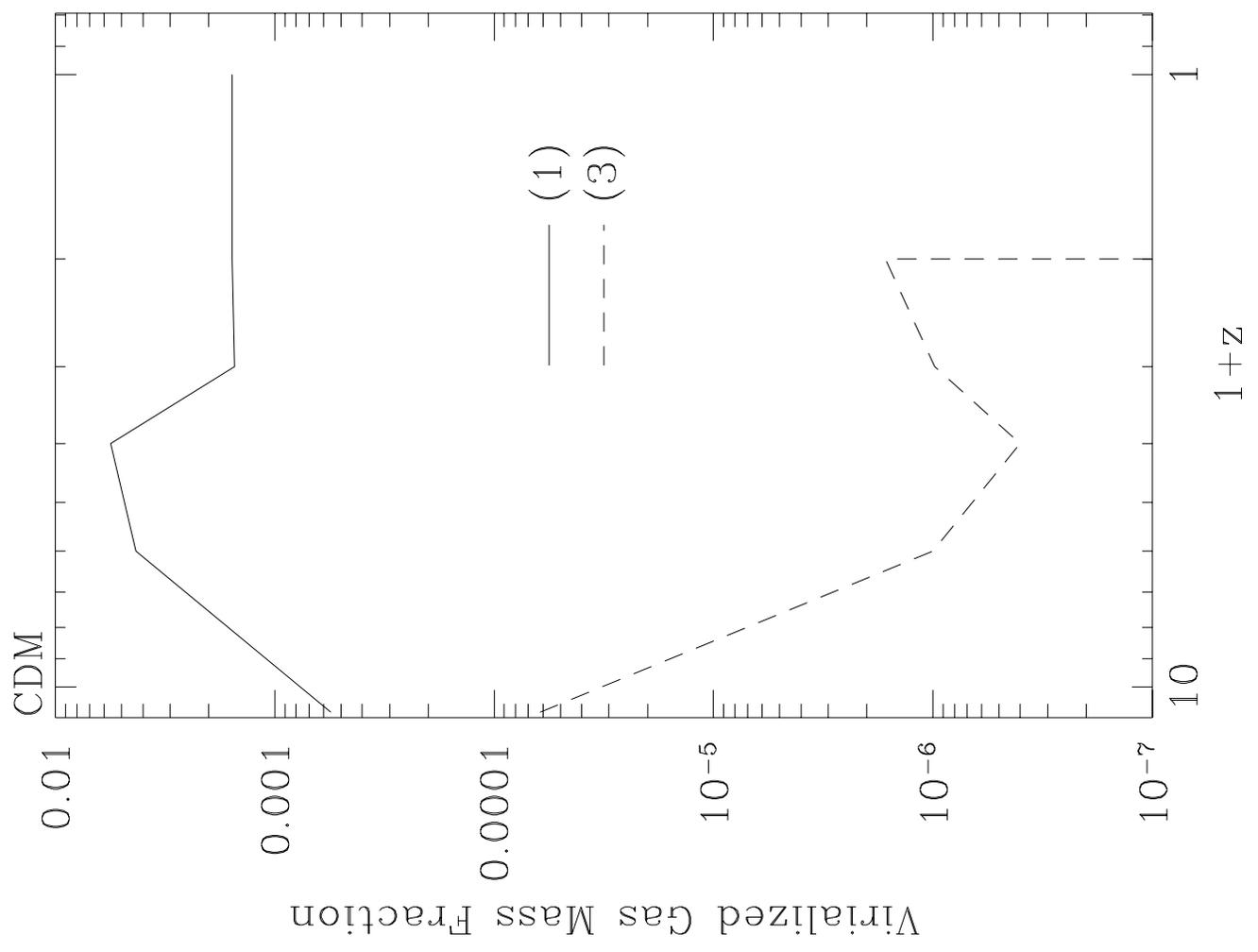

Figure 4c

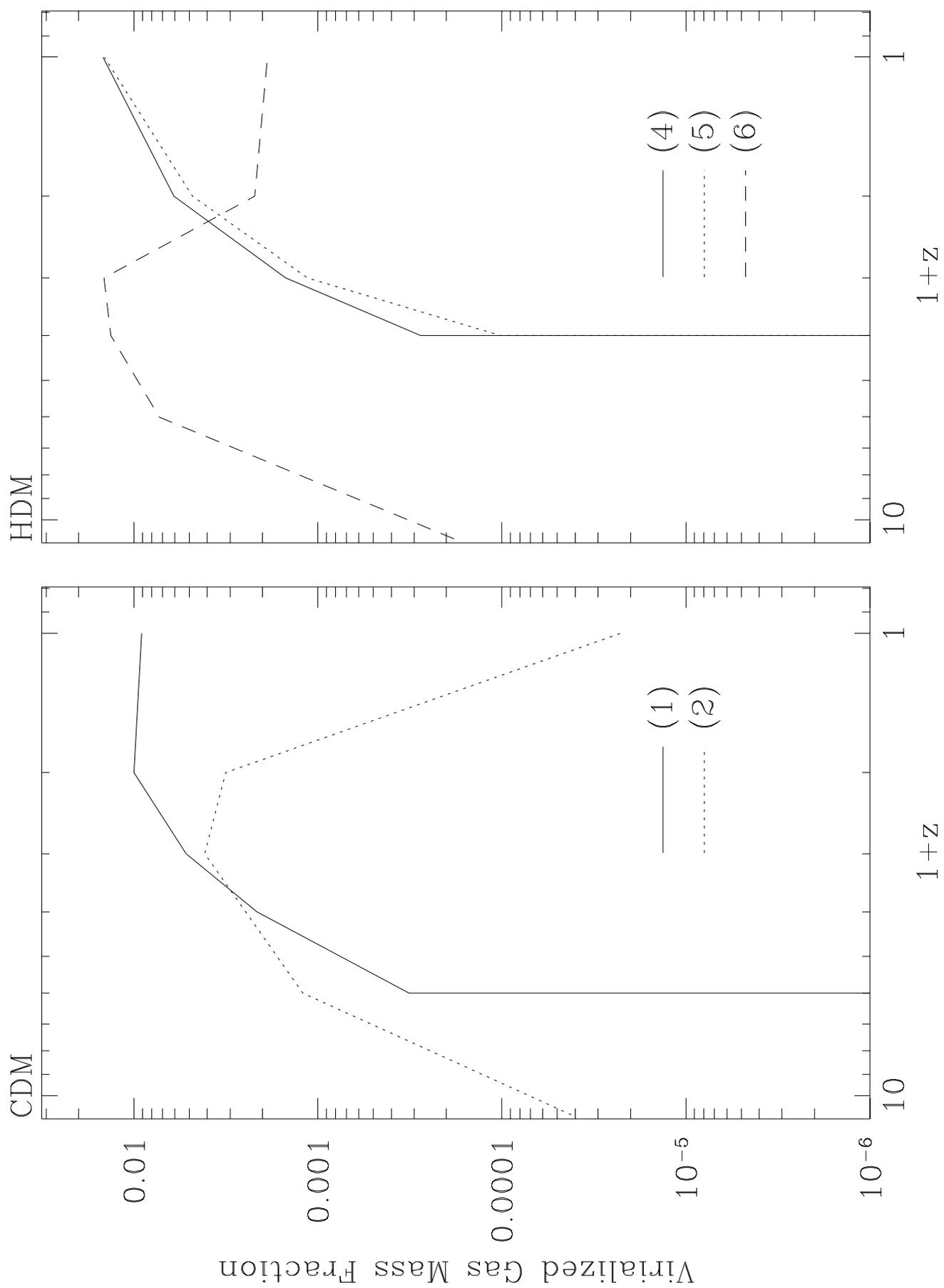
Figure 4b

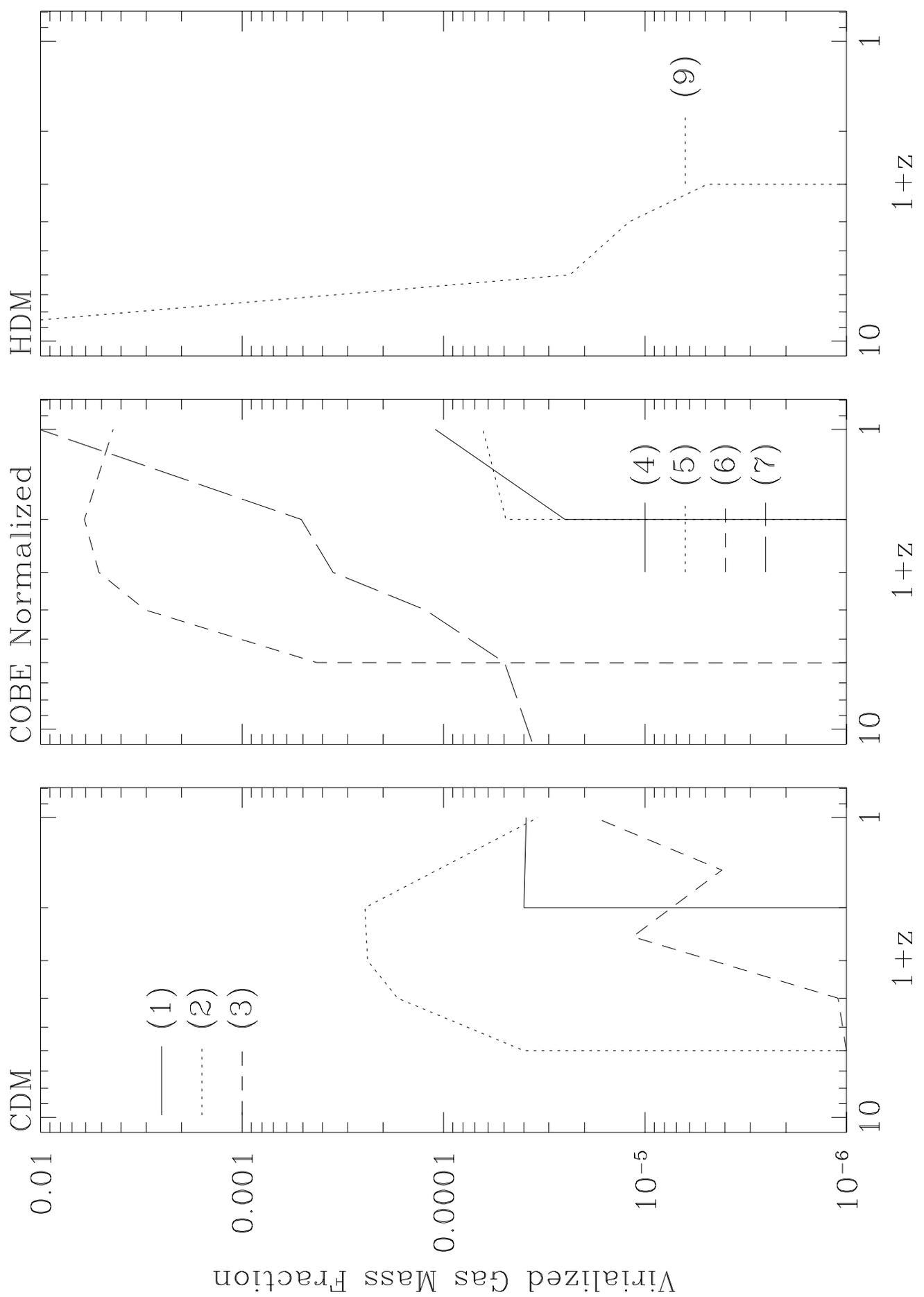

Figure 4a

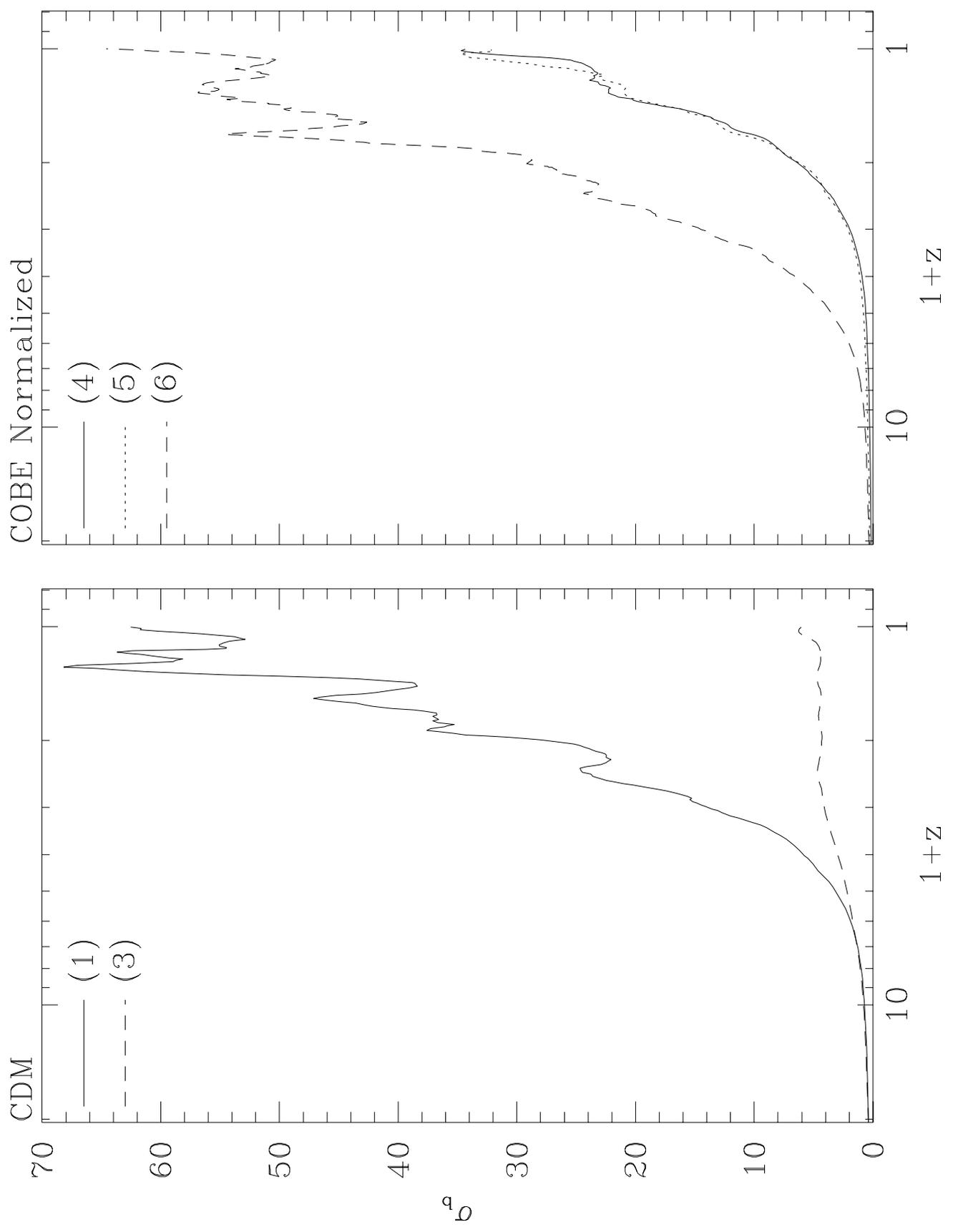

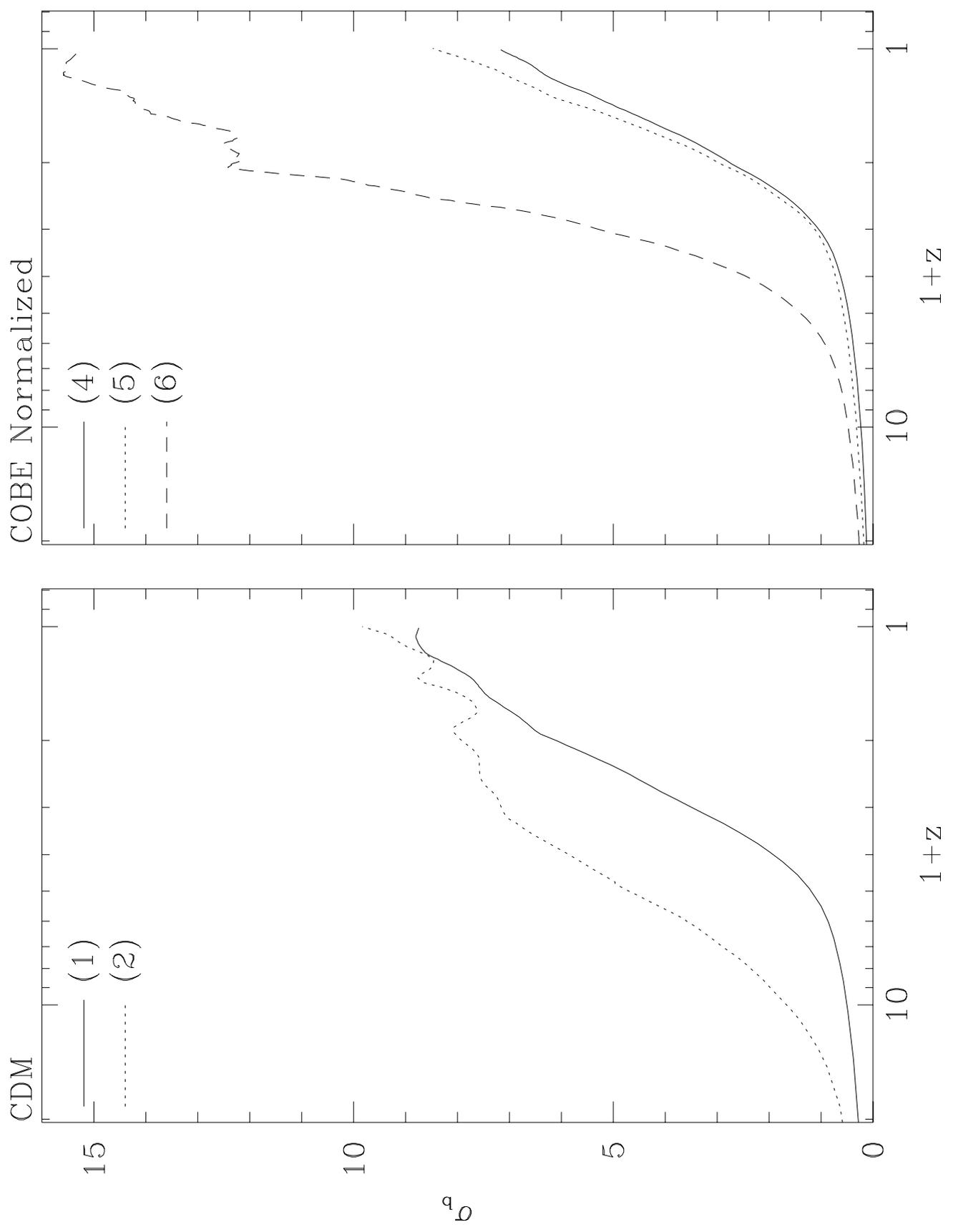
Figure 3b

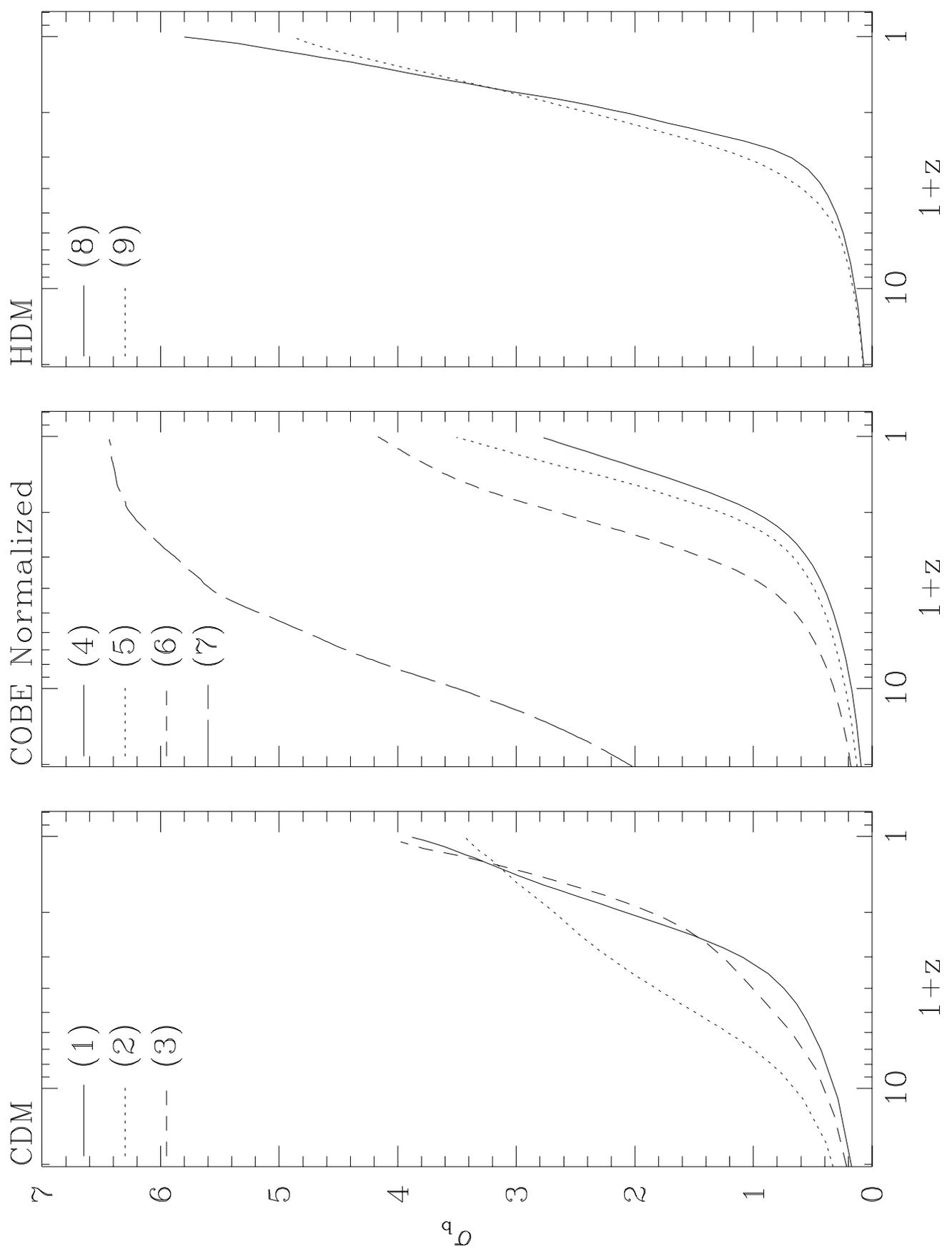

Figure 3a

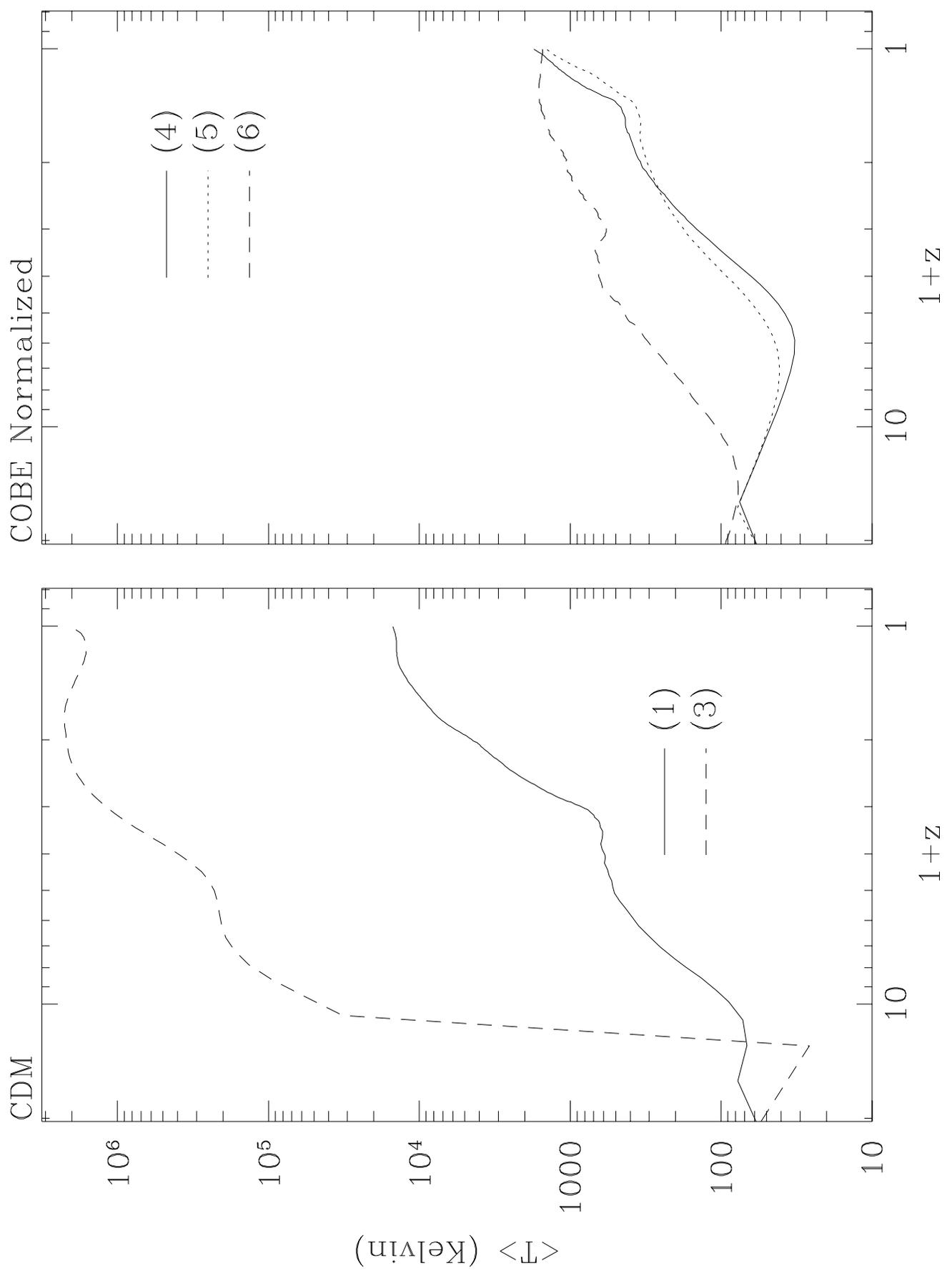
Figure 2c

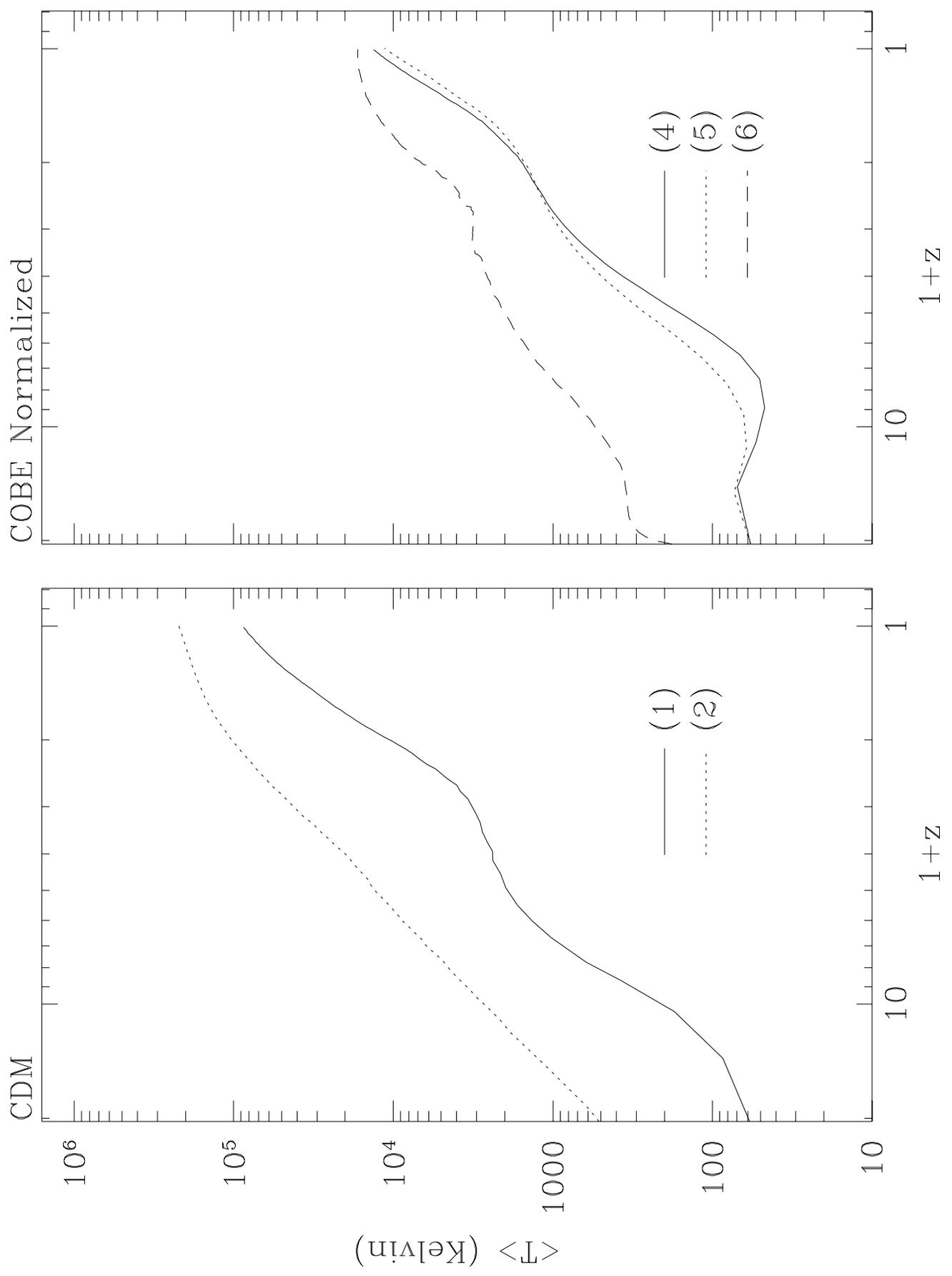

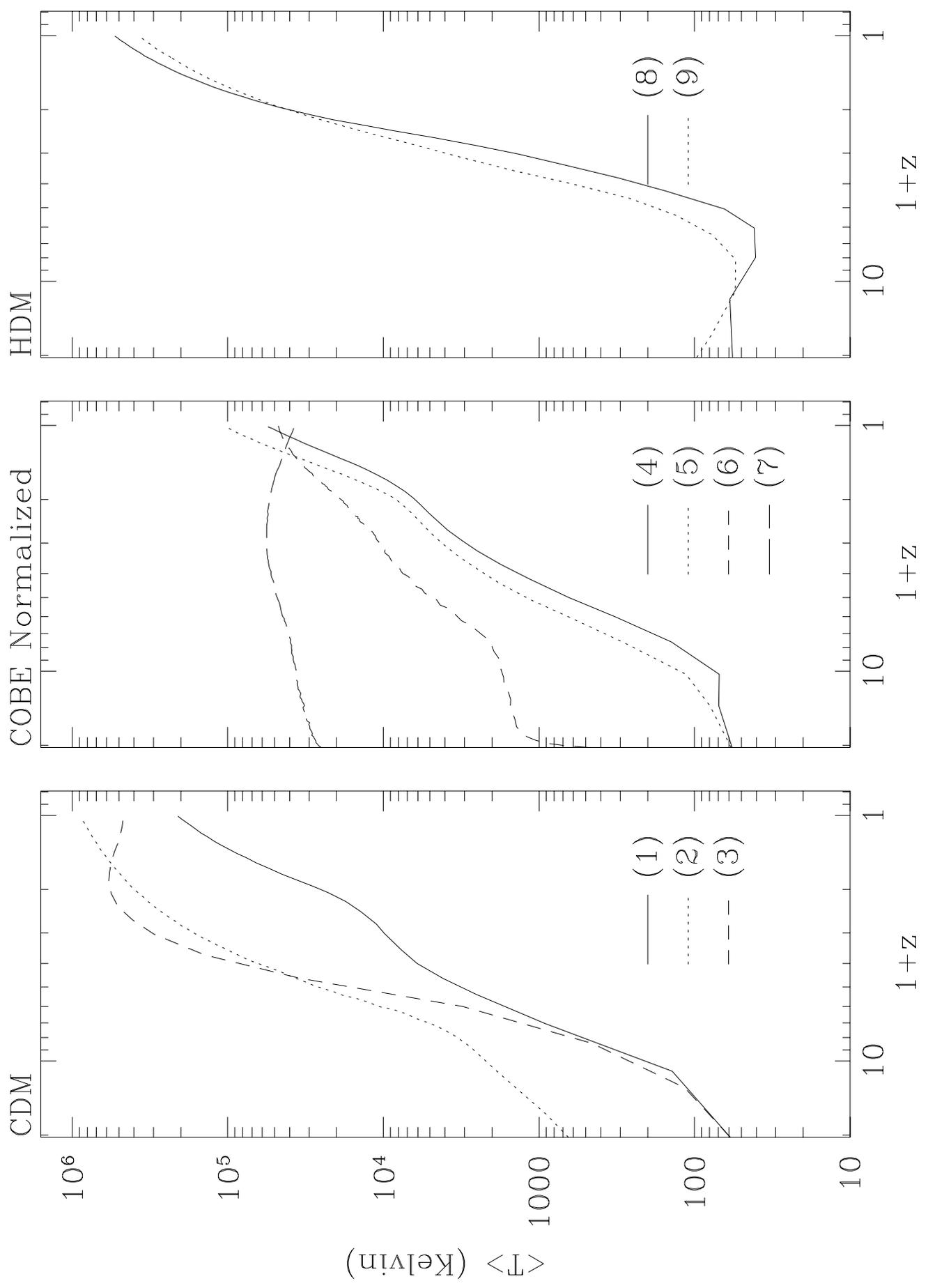

Figure 2a

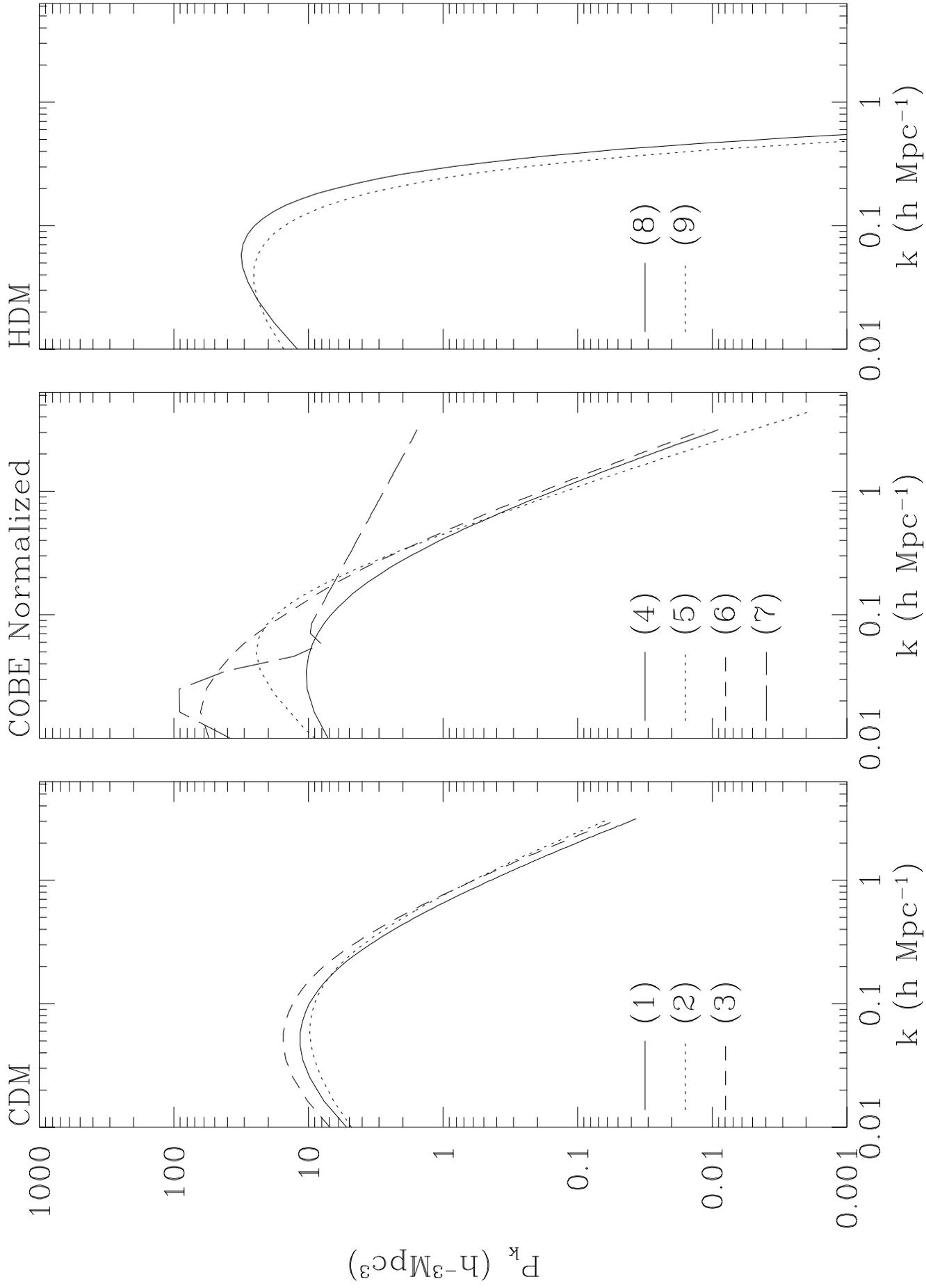

Figure 1